\documentclass{jfm}

\usepackage{lineno}
\usepackage{graphicx}
\usepackage{epstopdf, epsfig}
\usepackage{bm, amsmath}
\usepackage{verbatim}
\usepackage{color}
\usepackage{hyperref}
\usepackage{ulem}
\hypersetup{hypertex=true,
	colorlinks=true,
	linkcolor=blue,
	anchorcolor=blue,
	citecolor=blue}

\shorttitle{Drag reduction and flow statistics in Taylor-Couette turbulence with polymers}
\shortauthor{Y.-B. Zhang, Y. Fan, J. Su, H.-D. Xi and C. Sun}

\begin{document}
\title{Global drag reduction and local flow statistics in Taylor-Couette turbulence with dilute polymer additives}

\author{Yi-Bao Zhang\aff{1},
	Yaning Fan\aff{1},
	Jinghong Su\aff{1},
	Heng-Dong Xi\aff{2}\corresp{\email{hengdongxi@nwpu.edu.cn}},
	Chao Sun\aff{1,3}\corresp{\email{chaosun@tsinghua.edu.cn}}}

\affiliation{\aff{1}New Cornerstone Science Laboratory, Center for Combustion Energy, Key Laboratory for Thermal Science and Power Engineering of Ministry of Education, Department of Energy and Power Engineering, Tsinghua University, 100084 Beijing, China
\aff{2}Institute of Extreme Mechanics, School of Aeronautics, National Key Laboratory of Aircraft Configuration Design, Northwestern Polytechnical University, Xi'an 710072, China
\aff{3}Department of Engineering Mechanics, School of Aerospace Engineering, Tsinghua University, Beijing 100084, P. R. China}

\maketitle

\begin{abstract}
We present an experimental study on the drag reduction by polymers in Taylor-Couette turbulence at Reynolds numbers ($Re$) ranging from $4\times 10^3$ to $2.5\times 10^4$. In this $Re$ regime, the Taylor vortex is present and accounts for more than 50\% of the total angular velocity flux. Polyacrylamide polymers with two different average molecular weights are used. It is found that the drag reduction rate increases with polymer concentration and approaches the maximum drag reduction (MDR) limit. At MDR, the friction factor follows the $-0.58$ scaling, i.e., $C_f \sim Re^{-0.58}$, similar to channel/pipe flows. However, the drag reduction rate is about $20\%$ at MDR, which is much lower than that in channel/pipe flows at comparable $Re$. We also find that the Reynolds shear stress does not vanish and the slope of the mean azimuthal velocity profile in the logarithmic layer remains unchanged at MDR. These behaviours are reminiscent of the low drag reduction regime reported in channel flow (Warholic  \textit{et~al.}, \textit{Exp.~Fluids}, vol. {27}, issue 5, 1999, p. {461--472}). We reveal that the lower drag reduction rate originates from the fact that polymers strongly suppress the turbulent flow while only slightly weaken the mean Taylor vortex. We further show that polymers steady the velocity boundary layer and suppress the small-scale G\"{o}rtler vortices in the near-wall region. The former effect reduces the emission rate of both intense fast and slow plumes detached from the boundary layer, resulting in less flux transport from the inner cylinder to the outer one and reduces energy input into the bulk turbulent flow. Our results suggest that in turbulent flows, where secondary flow structures are statistically persistent and dominate the global transport properties of the system, the drag reduction efficiency of polymer additives is significantly diminished.

\end{abstract}


\section{Introduction}
The addition of long-chain flexible polymers with only a small amount into fluid flow can lead to fascinating phenomena very different from its Newtonian counterpart, among which the polymer drag reduction in wall-bounded turbulence is the most intriguing one \citep{White2008,Procaccia2008,Xi2019,Marchioli20212021009,saeed2023polymer}. Under specific conditions, the drag reduction rate can even reach $80\%$ \citep{virk1967toms}. Generally, the drag reduction rate increases with polymer concentration and approaches a maximum drag reduction (MDR) level, beyond which further increases in polymer concentration do not result in additional drag reduction. \cite{virk1970} summarized drag reduction experiments in pipe flow and found that at MDR, the friction factors from different datasets collapse and are limited by a unique asymptote: $C_f=0.59Re^{-0.58}$ for $Re\in \left[ 4\times10^3,\ 4\times 10^4\right]$. $Re$ is the Reynolds number based on the bulk velocity and pipe radius. Besides, at MDR, the mean streamwise velocity profile is also limited by an asymptote: $u^+=11.7{\rm ln} y^+-17$, where $u^+$ and $y^+$ are the mean streamwise velocity and wall normal distance normalized by the inner scale. The underlying mechanisms of drag reduction and MDR asymptote in channel/pipe flows remain an area of active research for decades. \cite{Xi2019} proposed the hibernating/active turbulence picture to explain the universality of the asymptote. At MDR, the active state is reduced by polymers and the hibernating state becomes the dominated dynamics. \cite{samanta2013elasto} found that when polymer solution exhibits high elasticity, it will bypass the Newtonian turbulence and experiences elasto-inertial instability with the flow friction coinciding with the MDR asymptote. Their results suggest that the asymptotic state is not related to ordinary turbulence but is associated with a new type of turbulent flow: elasto-inertial turbulence, where elasticity and inertia govern the flow dynamics \citep{choueiri2018exceeding}.

Taylor-Couette (TC) flow, the flow between two concentric cylinders, is also a canonic wall-bounded flow. TC flow has acted as a paradigmatic system of the physics of fluids, such as instabilities (see \cite{fardin2014} for a review) and high-Reynolds number turbulence (see \cite{grossmann2016} for a review). Moreover, TC flow has a close analogy with pipe flow \citep{eckhardt2000,eckhardt2007torque}. Compared with pipe/channel flow, TC flow has several advantages: (i) TC flow is a closed system, where the polymer concentration and flow properties (like temperature) can be well controlled. (ii) For closed systems, exact global balance relations between the driving and the dissipation can be derived.

TC flow laden with polymers is linearly unstable \citep{shaqfeh1996purely}. On average, polymer molecules orient and stretch in the streamwise direction. The resulting hoop stresses create extra pressure rendering the basic Couette flow unstable. At low $Re$ but high Weissenberg number, the elastic stress induces a spatially smooth and random in time turbulent flow, known as elastic turbulence \citep{groisman2004elastic,Groisman2000}. Elastic turbulence enhances the mixing but also significantly increases drag in the system \citep{Groisman2001,groisman2004elastic}. Polymers also alter the transition path of TC flow. In Newtonian fluid, with the increase of $Re$, the flow experiences in sequence the azimuthal Couette flow, axisymmetric Taylor vortex flow, wavy Taylor vortex flow, and turbulent Taylor vortex flow \citep{andereck1986flow}. In the viscoelastic fluid, many unique flow structures and patterns are reported, such as the diwhirls, rotating standing waves, disordered oscillations, flame patterns, and ribbons \citep{groisman1996couette,groisman1997solitary,dutcher2013effects,latrache2016defect,lacassagne2020vortex}. These patterns are also reproduced in numerical simulations coupled with the finitely extensible nonlinear elastic -Peterlin model \citep{thomas2006pattern}.

At higher $Re$ and high Weissenberg number, both inertia and elasticity dominate the dynamics. The transition to elasto-inertial turbulence in TC flow has been summarized in a recent review \citep{boulafentis2023experimental}. \cite{moazzen2023friction} studied the effect of polymers on the drag of the system in elasto-inertial turbulence. \cite{boulafentis2024coherent} measured the coherent structures and found that the dominated structures are dynamically independent solitary vortex pairs. These vortex pairs are sustained by the hoop stress and can emerge when moving sufficiently close. The spatial spectra of the intensity fluctuations in the elasto-inertial turbulence have been found to follow a power-law scaling with its exponent about $-7/3$ \citep{moazzen2023friction} and $-2.5$ \citep{boulafentis2024coherent}. These exponents are between $-3$, scaling for a smooth velocity field \citep{Groisman2000}, and $-5/3$, scaling for high $Re$ fully developed turbulence. Thus the degree of non-linearity of the underlying flow field in elasto-inertial turbulence is lower than that in inertial turbulence \citep{Zhang2021}.

At even higher $Re$, the inertial effect becomes dominant. Using laser-induced fluorescence, \cite{lee1995effect} studied the effect of polymers on the formation of small-scale G\"{o}rtler vortices near the inner cylinder. They observed that polymers suppress the formation of G\"{o}rtler vortices in the Reynolds number range of  $1483\leq Re \leq 30659$. \cite{rajappan2020cooperative} investigated the combined effect of superhydrophobic surfaces and polymer additives on the extent of drag reduction in TC flow. Polymers reduce about $20\%$ of the drag and hydrophobic surface can enhance the drag reduction effect of polymers in the Reynolds number range of  $15000\leq Re \leq 52000$. \cite{song2021direct} numerically investigated the drag enhancement effect at a constant Weissenberg number for $500\leq Re \leq 8000$. They divided the flow into two regimes: a low Reynolds number regime ($Re \leq 1000$), where elastic force dominates and large-scale structures contribute to momentum transport, and a high Reynolds number regime ($Re \geq 5000$), where inertial force dominates and small-scale near-wall structures govern the flow dynamics. \textcolor{black}{\cite{barbosa2022polymer} studied the effect of polymer concentration on the magnitude of drag reduction rate. They found that the MDR can be overpassed when polymer concentration is above the overlap concentration, and attribute this effect to polymer aggregation.} \textcolor{black}{In these studies, only the inner cylinder rotates. The Reynolds number is defined based on the velocity of the inner cylinder and the gap width between the two cylinders.}

The aforementioned studies mainly focus on the effect of polymers on the stability and transition of TC flow. These flows are characterized by low Reynolds numbers. At high $Re$, previous work employed visualization to study the effect of polymers on the G\"{o}rtler vortices \citep{lee1995effect} or torque measurement to study the drag modification \citep{rajappan2020cooperative,barbosa2022polymer}. Several fundamental questions remain: (i) whether the drag reduction behavior and mechanism in TC flow are similar to those in rectilinear flows (such as channel/pipe flow); (ii) the MDR asymptote in TC flows; (iii) the velocity field including mean velocity and turbulence statistics. \textcolor{black}{The first question addresses whether drag reduction behavior and mechanisms are universal across different flow systems.} In this study, we combine global torque measurement and local velocity field measurement to investigate polymer drag reduction in TC flow. The paper is organized as follows. In \S \ref{sec:meth}, the experimental set-up, the working fluid, the measurement technique, and the explored parameter space are described in detail. In \S \ref{sec:result}, we present our major results: including the friction-factor-Reynolds-number scaling, drag reduction contribution from the mean Taylor vortex and turbulence fluctuation, the effect of polymers on the energy distribution among different scales, and the velocity boundary layer. The paper ends with a summary and conclusion in \S \ref{sec:result}.

\section{Experimental setup and methodology}\label{sec:meth}

\subsection{Experimental setup and working fluid}
\begin{figure}
	\centerline
	{\includegraphics[width=0.8\columnwidth]{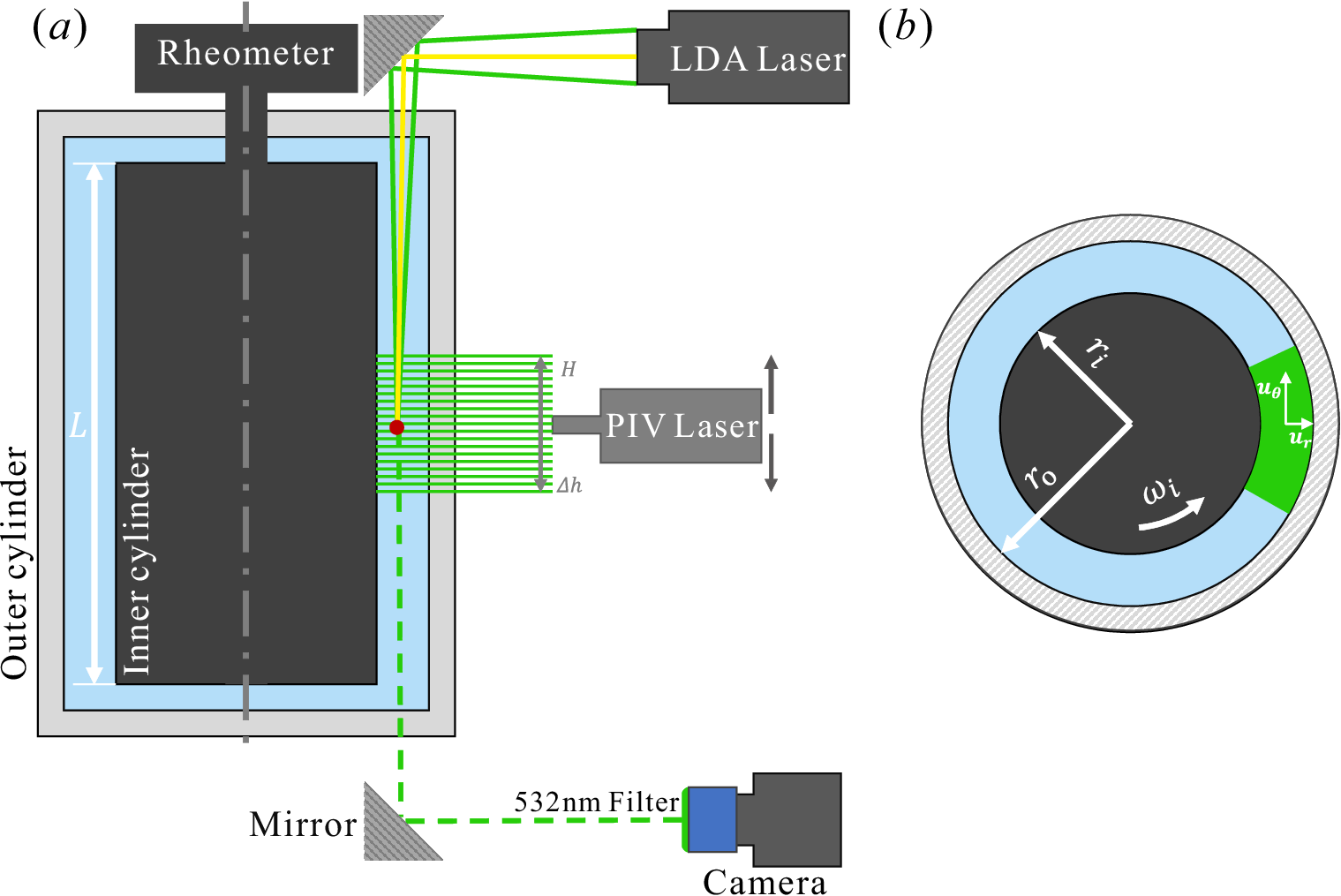}}
	\caption{(\textit{a}) A sketch of the experimental setup. The measurement techniques, including PIV and LDA, are also depicted. (\textit{b}) The two measured instantaneous velocity components are denoted as $u_\theta$ and $u_r$ in the azimuthal and radial directions, respectively.}
	\label{fig:setup}
\end{figure}

The TC system is constructed from a commercial rheometer (Discovery Hybrid Rheometer, TA Instruments) as shown in Fig. \ref{fig:setup}(\textit{a}). The inner cylinder, with a radius $r_i=25$ mm and height $L=100$ mm, is made of aluminium, which is anodized to form a black oxidation layer. This black surface can reduce unwanted reflection during the particle image velocimetry (PIV) and laser Doppler anemometer (LDA) measurements. The outer cylinder, with an inner radius $r_o=35$ mm and height 110 mm, is made of plexiglass, allowing for optical measurement. These two cylinders give a gap $d=r_o-r_i=10$ mm, a radius ratio $r_i/r_o=0.714$, and an aspect ratio $\Gamma=L/d=10$. The outer cylinder is enclosed inside a cubic tank (not shown in Fig. \ref{fig:setup}(\textit{a})). The gap between the outer cylinder and the tank can be circulated with water from a refrigerator to control the temperature of the working fluid, which is set to be $T=25\ ^\circ$C. A PT100 thermocouple is used to measure the temperature, with its variation remaining within $0.1$ K during the experiment.

The inner cylinder is connected to the rheometer and driven by its motor at a constant angular velocity $\omega_i$. The outer cylinder is fixed in our experiment. \textcolor{black}{The working fluid exerts frictional drag on the surface of the inner cylinder, which can be read from the TA Instruments Trios software as torque data. The torque resolution of the rheometer is 0.1 nN$\cdot$m.} This torque consists of two parts: one is related to the TC flow and the other is related to the von K\'{a}m\'{a}n flow formed between the ends of the two cylinders. The latter contribution can be estimated by using inner cylinder with different height, and the details can be found in previous studies \citep{yi2022physical,wang2022finite}. The torque contributed by the TC flow is denoted as $\tau$ in this study. For each solution, the torque measurement is repeated three times. Since the difference among repeated measurements is less than $1\%$, we omit the error bar of torque in the following.

The working fluid we used is a mixture of deionized water and glycerol, whose weight fraction is $w_g=25\%$. At $T=25\ ^\circ$C, the density and dynamic viscosity of the mixture are $\rho_s=1.058\times 10^3$ kg/m$^3$ and $\mu_s=1.79\times 10^{-3}$ Ns/m$^2$ \citep{wang2022finite}. Polyacrylamide polymers with two molecular weight $M_w$ are used: one with $M_w=5\times 10^6$ (Sigma-Aldrich) and is denoted as pam5e6; another with $M_w=2\times 10^7$ (Macklin), denoted as pam2e7. pam5e6 has been widely used in previous studies \citep{samanta2013elasto,choueiri2018exceeding}. The preparation protocol of the polymer solution can be found in our previous studies \citep{Zhang2021,zhang2022measured,peng2023}. The dynamic viscosity of the polymer solutions is measured by the Discovery Hybrid Rheometer equipped with a cone-plate geometry, whose diameter is 40 mm and angle is $2^\circ$. The shear viscosity $\mu_{\dot{\gamma}}$ as a function of shear rate $\dot{\gamma}$ for these two polymers is shown in the Appendix \ref{appA}. \textcolor{black}{The overlap concentration estimated from the zero-shear viscosity $\mu_0$ is $\phi^*\approx 680$ ppm (parts per million by weight) for pam5e6 and $\phi^*\approx 17$ ppm for pam2e7 (see Appendix \ref{appA})}. In this study, the used polymer concentration $\phi$ in TC is in the range of $0\leq \phi \leq 100$ ppm for pam5e6 and $0\leq \phi \leq 4$ ppm for pam2e7. Hence, the polymer solution is in the dilute regime. \textcolor{black}{The average shear rate in the gap of TC flow can be estimated to be $\omega_ir_i/d$, which is larger than 80 s$^{-1}$ in this study. When $\dot{\gamma}>80$, the shear-thinning effect of both fluids is found to be neglected. We therefore average $\mu_{\dot{\gamma}}$ over the range of $80<\dot{\gamma}<400$ to obtain an averaged viscosity $\mu$, and take this value as the viscosity of the working fluid (see Appendix \ref{appA} for more details).} In the dilute regime, the relaxation time $t_p$ of the polymer solution can be estimated from the Zimm model \citep{Ouellette2009,Zhang2021}, i.e., $t_p=N^{9/5}a^3\mu/(k_BT)$, where $N$ is the number of monomers, $a$ the length of one monomer, and $k_B$ the Boltzmann's constant.

\begin{figure}
	\centerline
	{\includegraphics[width=0.5\columnwidth]{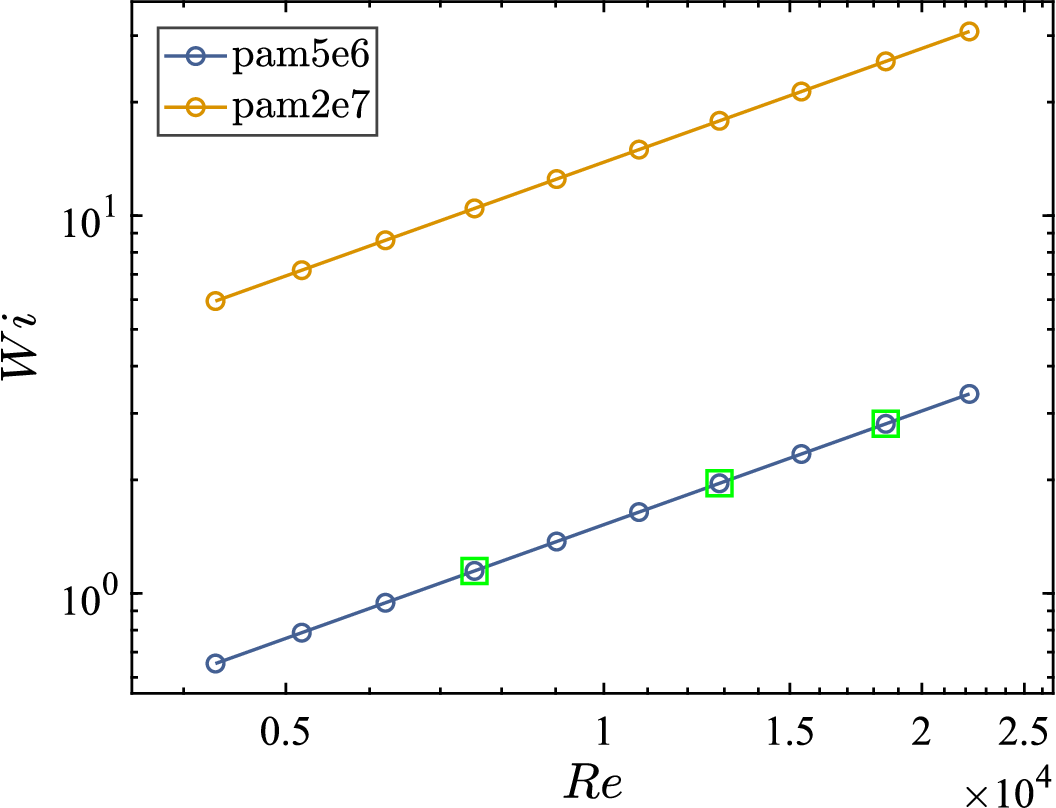}}
	\caption{Parameter space explored in this study. The green squares mark the parameters where we performed PIV and LDA measurements.}
	\label{fig:parameter}
\end{figure}

In TC turbulence, the control parameter is the Reynolds number, defined as 
\begin{equation}
Re=\omega_i r_i d/\nu,
\end{equation}
where $\nu=\mu/\rho_s$ is the kinematic viscosity of the solution; \textcolor{black}{or the Taylor number, defined as}
\begin{equation}
Ta=\frac{(1+r_i/r_o)^4}{64(r_i/r_o)^2}\frac{(r_o-r_i)^2(r_o+r_i)^2\omega^2_i}{\nu^2}.
\end{equation}
When using viscoelastic fluid, two additional parameters are needed: the polymer concentration $\phi$ and Weissenberg number defined as 
\begin{equation}
Wi=t_p\omega_ir_i/d.
\end{equation}
The response parameter of the system is the dimensionless torque given by 
\begin{equation}
G=\tau/(2\pi L \rho_s \nu^2), 
\end{equation}
which is related to the friction coefficient $C_f$ by the relation 
\begin{equation}
C_f=\left[ (1-r_i/r_o)^2/\pi \right]G/Re^2.
\end{equation}
The drag reduction rate by polymers is defined as 
\begin{equation}
DR=1-G_p/G_n, 
\end{equation}
where the subscripts $_p$ and $_n$ denote the viscoelastic case and Newtonian case, respectively. Please note that $G_p$ and $G_n$ are dimensionless torque measured at the same $Re$ instead of at the same $\omega_i$, namely we increase $\omega_i$ to keep $Re$ the same in the viscoelastic case. The parameter space explored in the present study is shown in Fig. \ref{fig:parameter}. The Reynolds number is in the range of $4\times 10^3<Re<2.5\times 10^4$, \textcolor{black}{and the corresponding Taylor number is in the range of $2.8\times 10^7<Ta<7.5\times 10^8$}. In this parameter range, the Taylor vortex is present and plays an important role in the flow dynamics \citep{ostilla2014boundary}. The $Wi$ for pam2e7 is nearly an order of magnitude larger than that for pam5e6. At high $Re$, it is observed that the torque slowly increases with time, signifying that the polymer molecules are degraded by the high shear near the inner cylinder. We therefore use the first one minute torque data after it becomes statistically steady to obtain the average $\tau$ for pam2e7.

In viscoelastic flow, the elastic number $El=Wi/Re$, which compares the elastic and inertial forces, is also an important parameter. In our study, $El=Wi/Re=t_p\nu/d^2$ is independent of $Re$. $El=1.5\times 10^{-4}$ and $1.3\times 10^{-3}$ for pam5e6 and pam2e7, respectively. According to \cite{song2023turbulent}, TC turbulence with $Re>4\times 10^3$ and $El<0.01$ is in the inertia-dominated regime.
 
\subsection{Methodology}

We employ PIV to measure the velocity in the $r-\theta$ plane, where $r$ and $\theta$ are in the radial and azimuthal directions, respectively. Due to the presence of Taylor vortices, the turbulence statistics depend on the axial height (we refer to the axial direction as $z$). The PIV measurements are conducted at 19 different heights which are separated by $\Delta h=1.5$ mm. Thus, the height of the measurements spans a length of $H=18\Delta h=27$ mm (Fig. \ref{fig:setup}(\textit{a})). The measurement length in unit of gap width is $H/d=2.7$, which is larger than the axial span of one pair of Taylor vortices observed in Newtonian turbulence \citep{froitzheim2019statistics}. Previous numerical simulations have reported that the Taylor vortex can be modified by polymers and its wavelength can either be decreased or increased \citep{song2021direct}. We therefore measure a larger height to test whether it is the case.

The seeding particles used are polystyrene with with an average diameter of 5 $\mu$m and density $1.05\times 10^3$ kg/m$^3$. The seeding particles are illuminated by laser sheet from a double-pulsed cavity laser (Beamtech Vlite200, 532 nm). The thickness of the laser sheet is $\approx 1$ mm. The laser is mounted on a traverse system (fly-opt PSTV50-S57) which allows us to precisely change the location of the laser sheet along the vertical direction. The images of the seeding particles are recorded by an HiSense Zyla CMOS camera (2560 × 2160 pixel, 16 bit). The PIV is operated on a double frame mode at a frame rate $f=4$ Hz, which is much smaller than \textcolor{black}{the characteristic frequency of the flow $\omega_i/(2\pi)$}, indicating that the measured velocity fields are independent with each other. At each height, we measure 500 velocity fields. During this time, the inner cylinder completes at least 1700 revolutions.

The captured image pairs are processed with a multi-pass algorithm. The interrogation window size is $64\times64$ and $32\times32$ in the first and second pass, respectively. In each pass, the overlap of the window is $50\%$. This process yields velocity fields in Cartesian coordinate with a spacing 0.11 mm between neighbour vectors. We then map the Cartesian velocity fields onto a polar grid using bilinear interpolation \citep{froitzheim2019statistics}. The measured velocity components in the radial and azimuthal directions are denoted as $u_r$ and $u_\theta$ (Fig. \ref{fig:setup}(\textit{b})), which are functions of $r$, $\theta$, $z$, and time $t$.  Since TC turbulence is statistically stationary and axisymmetric, the velocity field can be decomposed into its mean and fluctuation parts by the following relation, e.g., $u_\theta=\left\langle u_\theta \right\rangle_{t,\theta} + u^\prime_\theta$, where $\left\langle\ \right\rangle_{t,\theta}$ indicates average operation over time and in the $\theta$ direction. Here, we take $u_\theta$ as an example, similar operations can be done for $u_r$. The root mean square (rms) velocity of $u^\prime_\theta$ is defined as $\sigma_{u_\theta}=\sqrt{\left\langle u^{\prime 2}_\theta \right\rangle_{t,\theta,z}}$, where an additional average over the $z$ direction across one pair of Taylor vortices is applied. In the following, averaging in the $\theta$ direction is applied by default and therefore will be omitted in the subscript of the average operation.

We also perform LDA (TSI) measurement for its high temporal resolution. The LDA consists of two pairs of laser beams, and each pair measures one component of the velocity (see Fig. \ref{fig:setup}(\textit{a})). The LDA is mounted on a high-precision stage, which can move in the $x$, $y$, and $z$ directions independently. The measurements are done for two radial locations -- $(r-r_i)/d=0.06$ and $0.5$, and for three different heights -- vortex centre, outflow, and inflow. At the middle gap ($(r-r_i)/d=0.5$), we measure both $u_\theta$ and $u_r$. Near the inner cylinder ($(r-r_i)/d=0.06$), we can only measure $u_\theta$ due to the block of the laser beam by the inner cylinder. For each experiments we acquire $1\times 10^6$ data points. The PIV and LDA measurements are performed at three Reynolds numbers (marked as green squares in Fig. \ref{fig:parameter}) using the pam5e6 polymer due to its long-term stability.

\textcolor{black}{In TC flow, the torque of the system is related to the angular velocity flux $J^\omega$ by the relation $G=\nu^{-2}J^\omega$. $J^\omega$ is given by \citep{eckhardt2007torque}:}
\begin{equation} \label{eq:ang_vel_flux}
	J^\omega=r^3\left[ \left\langle u_r\omega \right\rangle_{t,z} -\nu\partial_r \left\langle \omega \right\rangle_{t,z} \right],
\end{equation}
\textcolor{black}{which is conserved along the radial direction. $\omega=u_\theta/r$ is the angular velocity. In the right hand side of Eq. \ref{eq:ang_vel_flux}, the first and second terms are the advective and viscous diffusion contributions, respectively. In analogy to the definition of the dimensionless heat flux in Rayleigh-B\'{e}nard flow, \cite{eckhardt2007torque} defined a Nusselt number as the ratio of $J^\omega$ and its value $J^\omega_{\rm lam}=2\nu r_i^2 r_o^2 \omega_i/(r_o^2-r_i^2)$ for the laminar case, i.e., $Nu_\omega=J^\omega/J^\omega_{\rm lam}$.}

\textcolor{black}{In the range of Reynolds number investigated here, the TC turbulence is a combination of turbulent Taylor vortices and background fluctuations. Thus, $r^3 \left\langle u_r\omega\right\rangle_{t,z}$ can be further decomposed into two parts \citep{brauckmann2013direct}, namely:}
\begin{equation} \label{eq:ang_vel_flux_adv}
	r^3 \left\langle u_r\omega\right\rangle_{t,z}=r^3 \langle \left\langle u_r\right\rangle_{t}\left\langle \omega \right\rangle_{t} \rangle_z+r^3 \left\langle u^\prime_r\omega^\prime \right\rangle_{t,z},
\end{equation}
\textcolor{black}{with the former one connected to the mean Taylor vortex motion and the latter one the turbulent fluctuations. The latter one can also be rewritten as $r^3 \left\langle u^\prime_r\omega^\prime \right\rangle_{t,z}=r^2 \left\langle u^\prime_r u_\theta^\prime \right\rangle_{t,z}$, which is the Reynolds stress contribution to the angular velocity flux. We denote the dimensionless form of these two contributions as $Nu_\omega^{tv}=r^3\left\langle u_r\right\rangle_{t,z}\left\langle \omega \right\rangle_{t,z}/J^\omega_{\rm lam}$ and $Nu_\omega^{rs}=r^2 \left\langle u^\prime_r u_\theta^\prime \right\rangle_{t,z}/J^\omega_{\rm lam}$.}

\begin{figure}
	\centerline
	{\includegraphics[width=0.8\columnwidth]{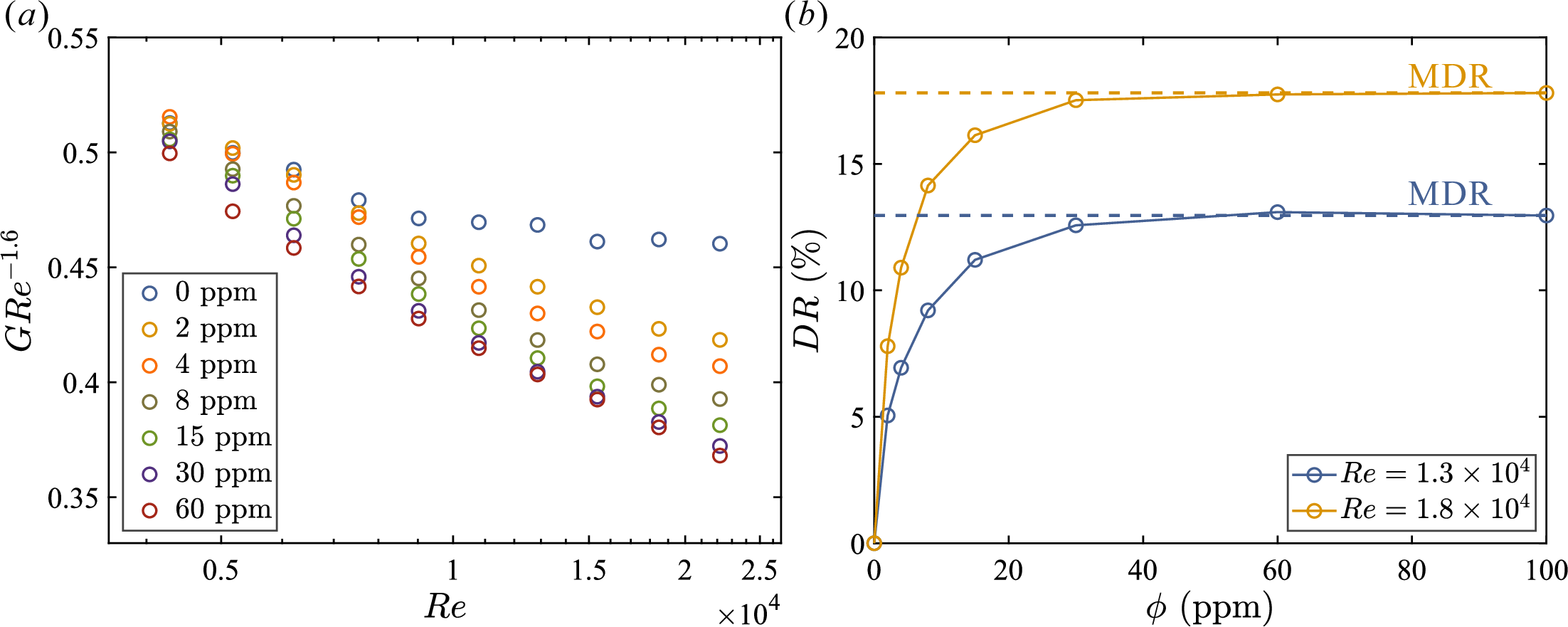}}
	\caption{(\textit{a}) Dimensionless torque compensated by the $1.6$ scaling, $GRe^{-1.6}$, as a function of the Reynolds number $Re$. (\textit{b}) Drag reduction rate, $DR$, for two typical $Re$. The horizontal dashed lines represent the maximum drag reduction rate. \textcolor{black}{The polymer we used here is pam5e6.}}
	\label{fig:torque}
\end{figure}

\begin{figure}
	\centerline
	{\includegraphics[width=0.8\columnwidth]{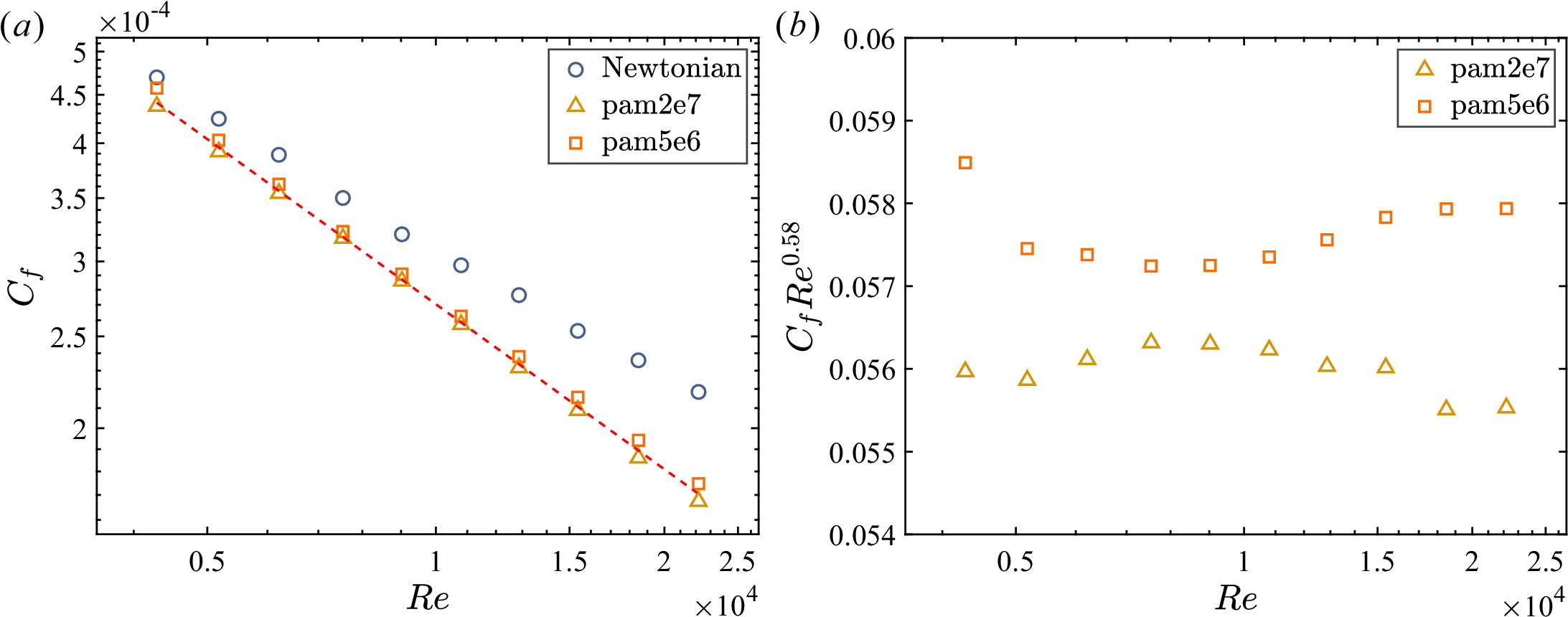}}
	\caption{(\textit{a}) Friction factor, $C_f$, for the Newtonian case and viscoelastic case under the maximum drag reduction condition. The dashed line has a slope of $-0.58$. (\textit{b}) $C_f$ compensated by the $Re^{-0.58}$ scaling.}
	\label{fig:mdr}
\end{figure}

\section{Results}\label{sec:result}
\subsection{Drag reduction characteristics}

Polymers reduce the drag of wall-bounded turbulence. In TC flow, this is reflected in the reduction of the dimensionless torque $G$ compared to the Newtonian case. \textcolor{black}{A comparison of $G$ in the Newtonian case between this study and \cite{lewis1999velocity} is made in Appendix \ref{appB}.} In Fig. \ref{fig:torque}(\textit{a}), we compare $G$ with increasing $\phi$ for different $Re$. To show the trend more clearly, $G$ is scaled by $Re^{-1.6}$ \citep{wang2022finite}. \textcolor{black}{In the range of $Re$ investigated here, the local scaling exponent for $G$ is not constant \citep{lewis1999velocity}. Therefore, $GRe^{-1.6}$ does not show a plateau in the Newtonian case ($\phi=0$ ppm) for the whole range of $Re$ investigated here. Instead, $GRe^{-1.6}$ is nearly constant for $Re$ in the range from $9\times10^3$ to $2.5\times 10^4$, in line with previous studies \citep{yi2022physical,wang2022finite}.} With the increase of $\phi$ and $Re$, $G$ for viscoelastic cases deviate more from their Newtonian counterparts, a feature consistent with what is reported in channel/pipe flows \citep{virk1975drag}. Additionally, the difference between neighboring $\phi$ values diminishes as $\phi$ increases, suggesting that the effects of polymers may saturate at high concentrations. To further strengthen this point, we show the drag reduction rate $DR$ as a function of $\phi$ for two typical $Re$ in Fig. \ref{fig:torque}(\textit{b}). $DR$ increases with $\phi$ and approaches an asymptote at high concentration, which is referred to as MDR \citep{Procaccia2008,White2008}.  

\textcolor{black}{In channel/pipe flow, the friction factor $C_f$ is generally used as the dimensionless drag of the flow system.} Here we also show $C_f$ of the Newtonian case and viscoelastic cases at the MDR for both pam5e6 and pam2e7 in Fig. \ref{fig:mdr}(\textit{a}). \textcolor{black}{At MDR, the two datasets nearly collapse and can be fitted with a power-law scaling $C_f\sim Re^{-(0.58\pm0.1)}$, which suggests that the friction factor at MDR in TC turbulence may also be universal, similar to what has been reported in pipe flows \citep{virk1970}.} In Fig. \ref{fig:mdr}(\textit{b}), $C_f$ is compensated by the $Re^{-0.58}$ scaling. $C_fRe^{0.58}$ exhibits plateau-like behaviour for both pam5e6 and pam2e7, demonstrating the robustness of this scaling. For the two polymers used in this study, the $Wi$ for pam2e7 is about ten times of the $Wi$ for pam5e6 (see Fig. \ref{fig:parameter}). However, the drag reduction rate differs by only about $3\%$ at MDR. In experiments, the MDR asymptote generally is approached by increasing the polymer concentration; however it is approached by the increase of Weissenberg number while keeping the concentration constant in numerical simulations \citep{Procaccia2008,Xi2019}. The critical condition under which the MDR limit is realized is a function of both $Wi$ and $\phi$, and there is no systematic study on this up to now, due to the large parameter space. Studies in this direction would have important implication on practical application of polymer drag reduction.

\begin{figure}
	\centerline
	{\includegraphics[width=0.5\columnwidth]{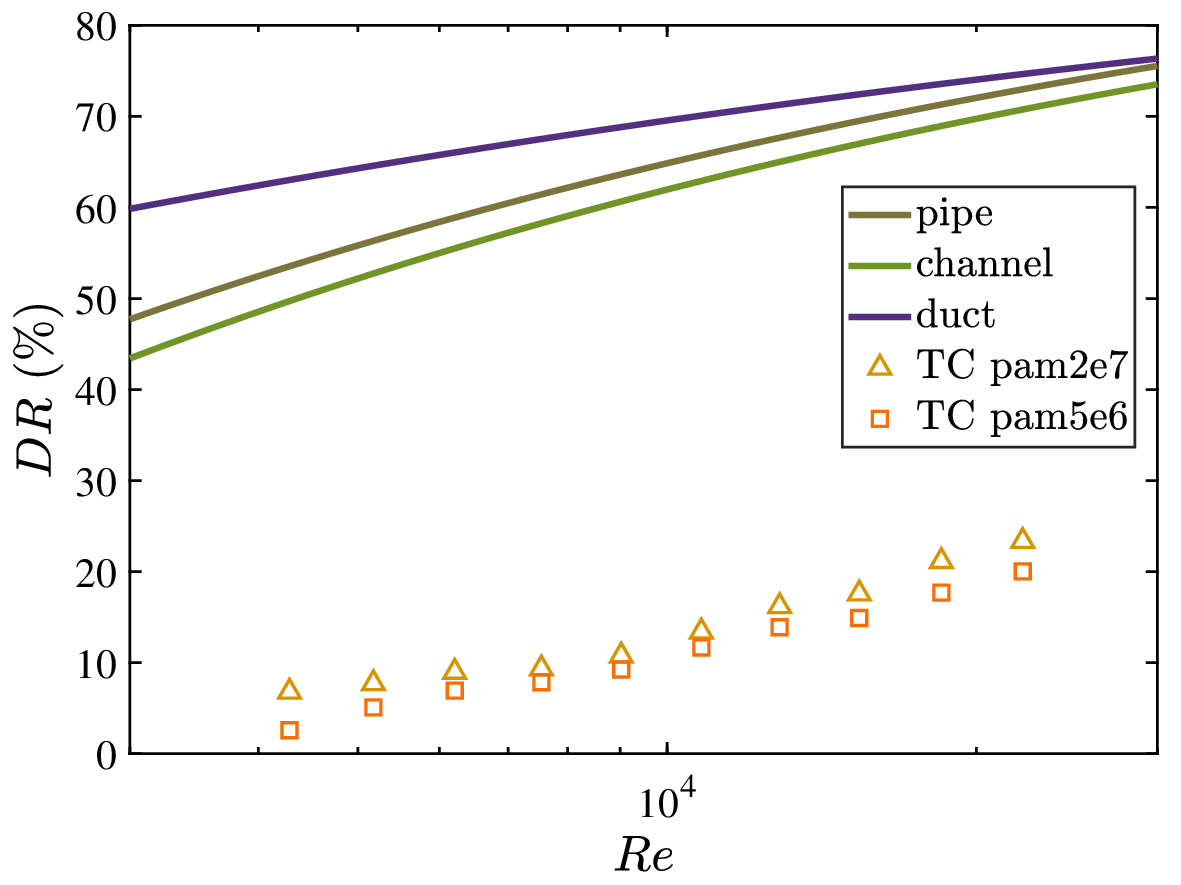}}
	\caption{Drag reduction rate at MDR for rectilinear flows (including pipe, channel, and duct flows) and Taylor-Couette turbulence.}
	\label{fig:flow_mdr}
\end{figure}

\cite{virk1970} summarized previous studies about drag reduction by polymers in pipe flow, and found that $C_f$ is limited by an asymptote described by the experimental correlation: $C_f=0.59 Re^{-0.58}$ for $4\times10^3<Re<4\times 10^4$. While the scaling exponents are close in TC and pipe flows, the extent of drag reduction in TC is much smaller. For comparison, we compute the drag reduction rate for typical rectilinear flows, including pipe, channel and duct flows, under the MDR condition. Here, the drag reduction rate can be defined as $DR=1-C_{f,p}/C_{f,n}$. The friction factors of the rectilinear flows for the Newtonian and viscoelastic cases are from \cite{owolabi2017turbulent}. The drag reduction rate, $DR$, are compared in Fig. \ref{fig:flow_mdr}. For rectilinear flows, $DR$ can reach about $60\%$ at $Re\approx 1\times 10^4$ \citep{owolabi2017turbulent}. However, in TC flow, the maximum drag reduction is about $23\%$ at $Re=2.2\times10^4$. The level of drag reduction rate in our study is consistent with the results reported in numerical simulation \citep{lin2022high} and in experiments \citep{rajappan2020cooperative,barbosa2022polymer}. In their study, \cite{rajappan2020cooperative} used two kinds of polymers: polyacrylamide and polyethylene oxide. 

\subsection{Drag reduction mechanism}

\textcolor{black}{In TC turbulence, the viscous diffusion dominates in the near wall region while the advective contribution dominates in the bulk region of the gap \citep{brauckmann2013direct}. Unfortunately, the diffusion term cannot be accurately measured due to the low spatial resolution of our PIV. For TC flow laden with polymers, along with the advective and viscous diffusion contributions in Eq. \ref{eq:ang_vel_flux}, polymer stress also contributes to the angular velocity flux \citep{song2021direct}. However, this contribution cannot be directly measured experimentally. In channel and TC flows, the polymeric contribution to the drag is positive \citep{min2003drag,song2021direct}. One thus may conjecture that the advective contribution will be reduced by polymers in the bulk region. In this section, we try to reveal the underlying mechanism of the low drag reduction rate observed in TC turbulence by investigating the two parts of the advective contribution according to Eq. \ref{eq:ang_vel_flux_adv}: the Reynolds shear stress and the Taylor vortex.}

\begin{figure}
	\centerline
	{\includegraphics[width=0.8\columnwidth]{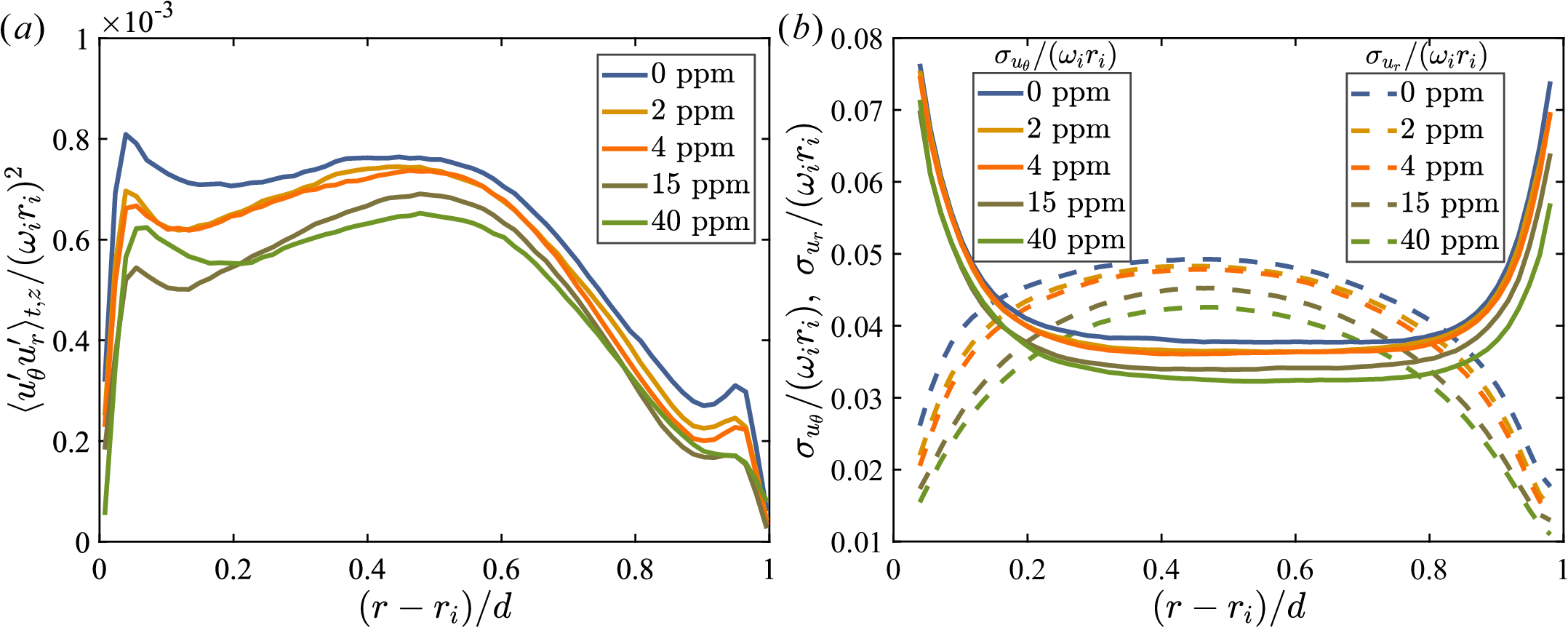}}
	\caption{(\textit{a}) Radial dependence of the Reynolds shear stress normalized by the velocity of the inner cylinder, i.e., $\langle u^\prime_\theta u^\prime_r\rangle_{t,z}/(\omega_i r_i)^2$. (\textit{b}) Radial dependence of the azimuthal (solid lines) and radial (dashed lines) rms velocity. $Re=1.3\times 10^4$.}
	\label{fig:rs}
\end{figure}

\textcolor{black}{We first study the effect of polymers on the Reynolds shear stress. The radial dependence of the Reynolds shear stress, normalized by the inner cylinder velocity $\omega_i r_i$, is shown in Fig. \ref{fig:rs}(\textit{a})}. In the near wall region, $\left\langle u^\prime_r u_\theta^\prime \right\rangle_{t,z}$ exhibits local maximum and minimum with their intensity higher near the inner cylinder than near the outer one, similar to the streamwise vorticity fluctuations observed by \cite{song2021direct}. In wall bounded turbulence, the local maximum in the Reynolds shear stress is found to be closely connected with the near-wall quasi-streamwise vortices \citep{jimenez2018coherent}. In TC turbulence, the near-wall coherent structures are dominated by G\"{o}rtler vortices \citep{dong2007direct}. \cite{dong2007direct} also found that the intensity of G\"{o}rtler vortices near the inner cylinder is stronger than near the outer one. The peaks near both cylinders decrease with the addition of polymers, suggesting that the G\"{o}rtler vortices are suppressed by polymers, a phenomenon that was directly observed by laser induced fluorescence experiments \citep{lee1995effect}. This is also similar to the effects of polymers on channel flow, where polymers suppress the near-wall quasi-streamwise vortices and lead to a reduction of Reynolds shear stress \citep{kim2008dynamics,min2003drag}.

\textcolor{black}{In Fig. \ref{fig:rs}(\textit{a}), it can be seen that the Reynolds shear stress is suppressed by polymers, and the suppression effect strengthens as the polymer concentration increases. The decrease in Reynolds shear stress implies that the turbulent flow contributes less drag to TC flow laden with polymers according to Eqs. \ref{eq:ang_vel_flux} and \ref{eq:ang_vel_flux_adv}. We also note that at MDR, the Reynolds shear stress in the viscoelastic case is smaller than in the Newtonian case but still of the same order. This behavior differs from what has been reported in channel/pipe flows, where the Reynolds shear stress is an order of magnitude smaller than in the Newtonian case and nearly vanishes at MDR \citep{warholic1999influence,choueiri2018exceeding}}. That is to say that the turbulent contribution to the drag can be ignored in channel/pipe flows when approaching the MDR asymptote.

The rms velocity in the azimuthal and radial directions, $\sigma_{u_\theta}/(\omega_i r_i)$ and $\sigma_{u_r}/(\omega_i r_i)$, are presented in Fig. \ref{fig:rs}(\textit{b}). It can be observed that $\sigma_{u_\theta}/(\omega_i r_i)$ and $\sigma_{u_r}/(\omega_i r_i)$ are reduced across the gap by the polymers. And the reduction effect becomes stronger with the increase of polymer concentration.

\begin{figure}
	\centerline
	{\includegraphics[width=0.5\columnwidth]{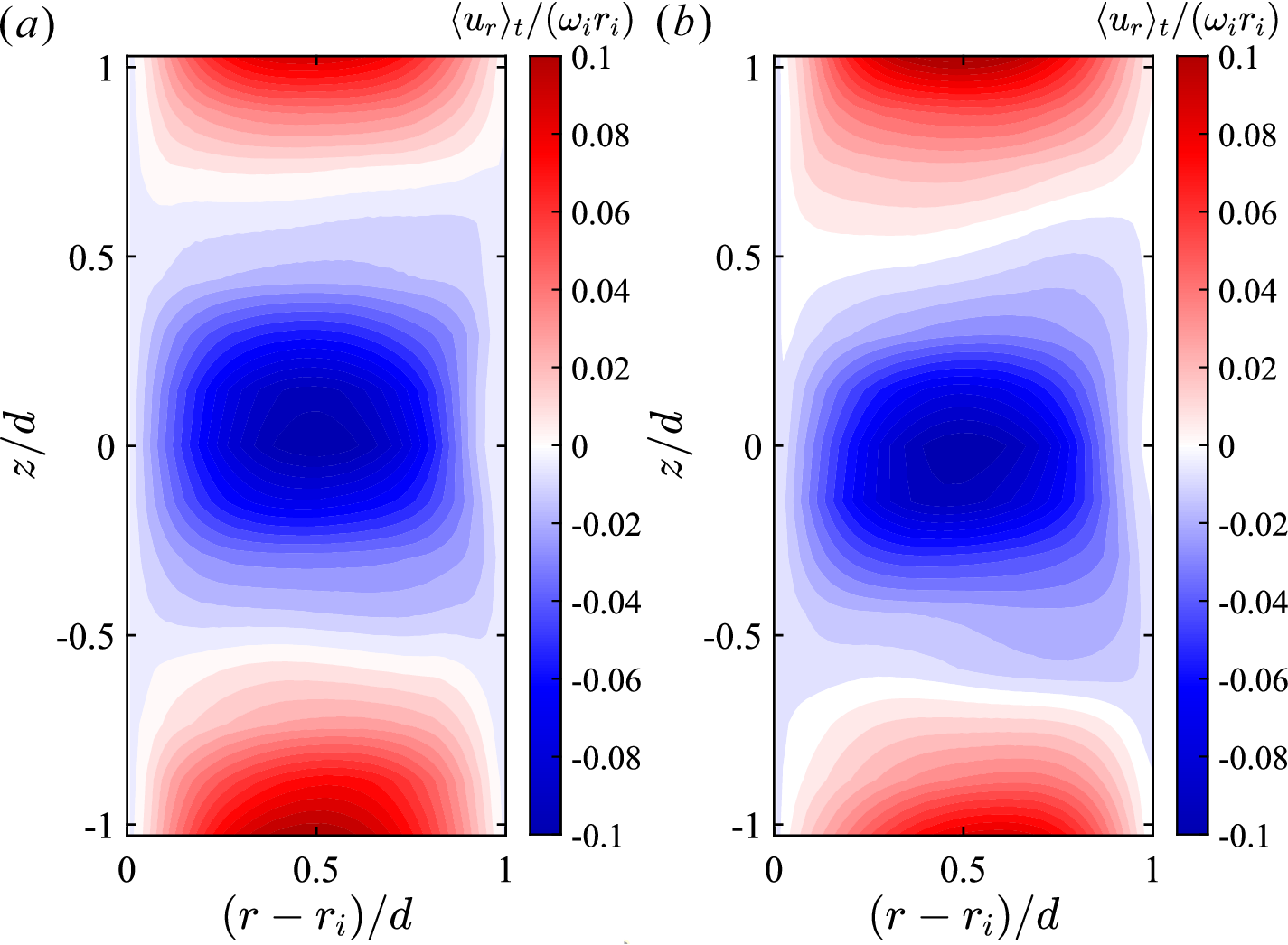}}
	\caption{Contours of the mean radial velocity, $\langle u_r \rangle_t/(\omega_i r_i)$, for the Newtonian case (\textit{a}) and viscoelastic case at $\phi=40$ ppm (\textit{b}), respectively. $Re=1.3\times 10^4$.}
	\label{fig:tv}
\end{figure}

\textcolor{black}{We then study the effect of polymers on the Taylor vortex.} As $Re$ increases, the turbulent fluctuation grows with the strength of Taylor vortex diminishing when only the inner cylinder rotating \citep{ostilla2014boundary}. \cite{brauckmann2013direct} found that the Taylor vortex contributes to about half of the total angular velocity flux in the bulk region for $Re\approx 2\times 10^4$. \textcolor{black}{For smaller $Re$, this contribution becomes even larger since the Taylor vortex occupies a larger fraction of the total energy in the flow.} In Fig. \ref{fig:tv}, we compare the contour of the mean radial velocity from the Newtonian case (\textit{a}) and viscoelastic case at $\phi=40$ ppm (\textit{b}) at $Re = 1.3\times 10^4$. We observe that the wavelength and pattern of the Taylor vortex are nearly the same for these two cases. In the parameter space of this study, the inertia dominates the flow dynamics \citep{song2023turbulent}, we thus can expect that the Taylor vortex, which is an inertial effect, would hardly be modified by polymers. This result is consistent with the numerical simulation of \cite{song2021direct}, which reported that increasing fluid inertia (while keeping $Wi$ constant) hinders the elastic effects.

\begin{figure}
	\centerline
	{\includegraphics[width=0.6\columnwidth]{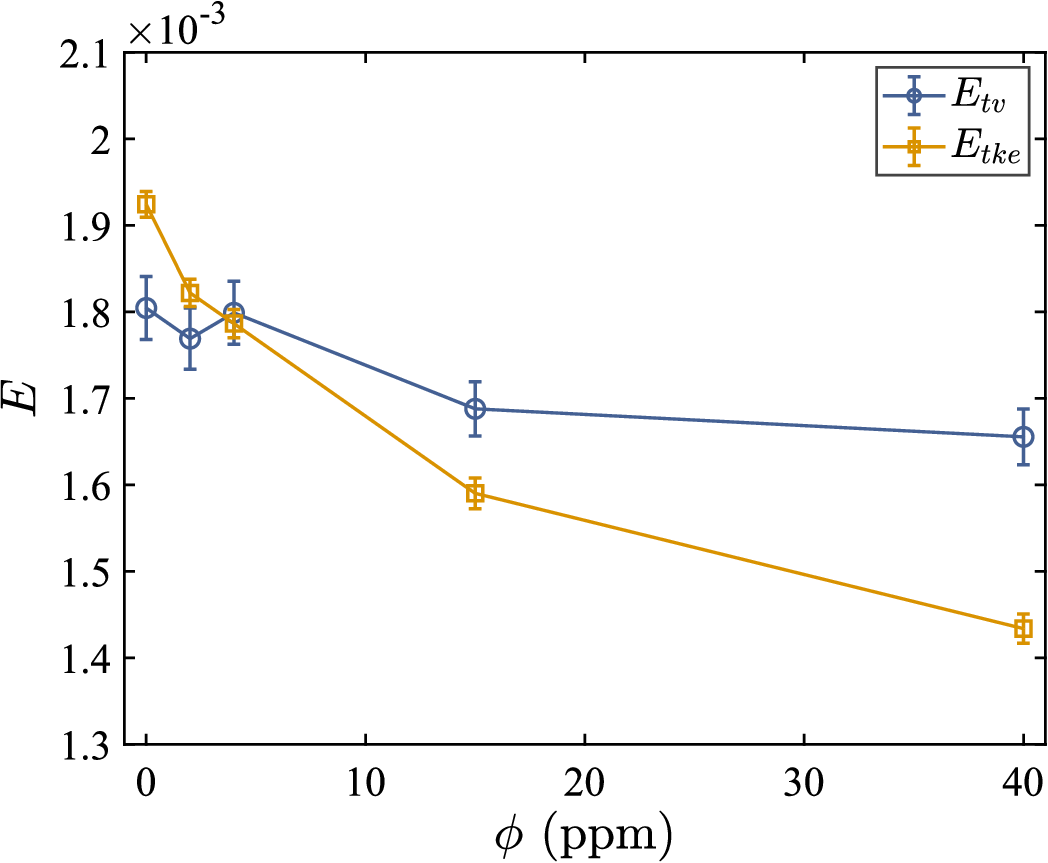}}
	\caption{The energy of the Taylor vortex, $E_{tv}$, and the turbulent kinetic energy, $E_{tke}$, as a function of polymer concentration. $Re=1.3\times 10^4$.}
	\label{fig:energy}
\end{figure}

While the wavelength and pattern of the Taylor vortex are nearly not altered by polymers, its strength is weakened. To quantify this, we adopt the method proposed in \cite{Froitzheim2017}. The kinetic energy of the flow averaged over time and space reads
\begin{equation}
	E(r)=\frac{1}{2}\left\lbrace \left\langle \left\langle u_\theta \right\rangle_t \right\rangle_{z}^2+\left\langle \left\langle u_r \right\rangle_t \right\rangle_{z}^2 +\left\langle \left\langle u_z \right\rangle_t \right\rangle_{z}^2 +\sigma^2_{u_\theta}+\sigma^2_{u_r}+\sigma^2_{u_z}   \right\rbrace.
\end{equation} \label{eq:ke}
The first term is the angular base flow, the main contribution to $E(r)$. To reveal the secondary flow and the turbulent fluctuations, we neglect this energy portion. Since the axial velocity is not measured, we only use the central region of the velocity to define the energy, with the assumption that the axial velocity of an idealized turbulent Taylor vortex vanishes in the centre of the gap \citep{Froitzheim2017}. To be specific, the energy related to the mean Taylor vortex and turbulent kinetic energy are given by:
\begin{gather}
	E_{tv}=\frac{1}{2}\left\langle \left\langle u_r ((r-r_i)/d=0.5) \right\rangle_t \right\rangle_{z}^2/(\omega_i r_i)^2,\\ E_{tke}=\frac{1}{2}\left\langle \sigma^2_{u_\theta}((r-r_i)/d=0.5)+\sigma^2_{u_r}((r-r_i)/d=0.5) \right\rangle_{z}^2/(\omega_i r_i)^2,
\end{gather}
and they are normalized by $(\omega_i r_i)^2$.

$E_{tv}$ and $E_{tke}$ as a function of polymer concentration are shown in Fig. \ref{fig:energy}. It is found that, while $E_{tke}$ is strongly depressed by polymers with the increase of polymer concentration, $E_{tv}$ is weakened by a less degree when $\phi>4$ ppm. We conclude that the major contribution to drag reduction by polymers comes from the reduction in turbulence, with the weakened Taylor vortex playing a minor role in the current parameter regime of TC turbulence \textcolor{black}{based on Eq. \ref{eq:ang_vel_flux_adv}}. The lower drag reduction rate in the current parameter regime of TC turbulence compared to channel/pipe/duct flows can be understood as follows: the angular velocity flux from the mean Taylor vortex motion is the dominating part of the system's drag, which is larger than $50\%$ in the range of $Re$ investigated in this study \citep{brauckmann2013direct}. However, polymers only have a minor effect on the Taylor vortex. Besides, due to the turbulent nature of the Taylor vortex, the Reynolds shear stress cannot be reduced to vanish at MDR as observed in Fig. \ref{fig:rs}(\textit{a}).

\begin{figure}
	\centerline
	{\includegraphics[width=0.8\columnwidth]{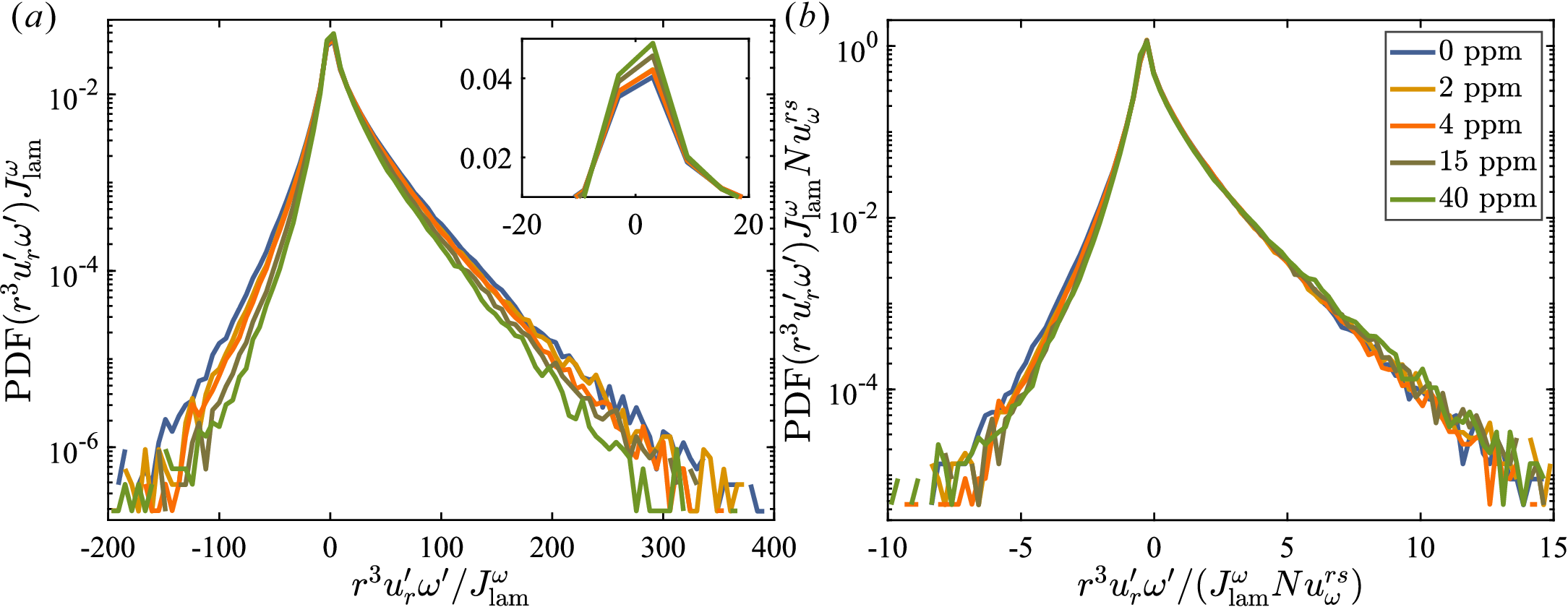}}
	\caption{(\textit{a}) Probability density functions of the angular velocity flux contributed by the Reynolds shear stress, $r^3 u^\prime_r \omega^\prime$, which is nondimensionalized by the value for laminar flow $J^\omega_{\rm lam}$. \textcolor{black}{Inset: an enlarged view of the main plot near its peak in a linear-linear scale.} (\textit{b}) Probability density functions of $r^3 u^\prime_r \omega^\prime/J^\omega_{\rm lam}$ scaled by its mean value, i.e., $r^2 u^\prime_r u_\theta^\prime/(J^\omega_{\rm lam}Nu_\omega^{rs})$. The probability density functions are calculated from data sampled on cylinder surface with $(r-r_i)/d=0.5$ at $Re=1.3\times 10^4$.}
	\label{fig:pdf_nu}
\end{figure}

To understand how polymers suppress turbulence and weaken the Taylor vortex, we delve into the probability density functions (PDFs) of the angular velocity flux from the Reynolds shear stress, i.e., $r^3 u^\prime_r \omega^\prime$, which is shown in Fig. \ref{fig:pdf_nu}(\textit{a}). Both positive and negative tails of the PDFs shrink with the increase of polymer concentration. When $r^3 u^\prime_r \omega^\prime$ are scaled by their mean value, as shown in Fig. \ref{fig:pdf_nu}(\textit{b}), \textcolor{black}{their positive tails nearly collapse with each other}. However, the negative tails exhibit a slight shrinkage. In TC turbulence, the angular velocity plumes detached from the velocity boundary layer are observed to be the underlying structure of efficient angular velocity carriers \citep{brauckmann2013direct,ostilla2014boundary,froitzheim2019statistics}. The shrinkage of the PDFs in their tails may suggest that the angular velocity plumes are modified by polymers.


\begin{figure}
	\centerline
	{\includegraphics[width=0.8\columnwidth]{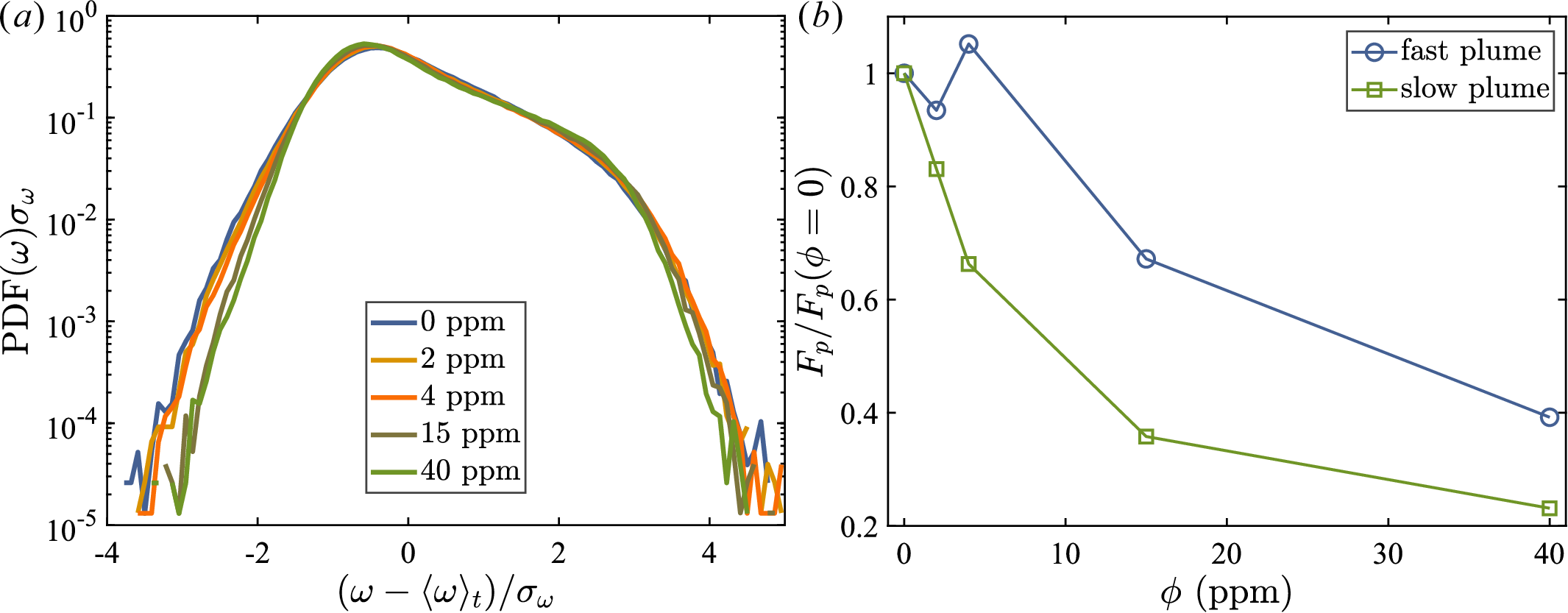}}
	\caption{(\textit{a}) Probability density functions of the angular velocity, $\omega$, sampled on cylinder surface $(r-r_i)/d=0.03$. $\omega$ is normalized by its mean value and standard deviation in such a way that the integration of the curve with respect to the abscissa is $1$. (\textit{b}) Plume fraction $F_p$ normalized by its value from the Newtonian case $F_p(\phi=0)$. $Re=1.3\times 10^4$.}
	\label{fig:pdf_uomega}
\end{figure}


\textcolor{black}{At $Re=1.3\times 10^4$, the average slope boundary layer thickness in the gap can be estimated as $\lambda_\omega=d/(2Nu_\omega)\approx0.37$ mm, following the theory in \cite{eckhardt2007torque}.} We plot in Fig. \ref{fig:pdf_uomega}(\textit{a}) the PDFs of the normalized azimuthal velocity by their mean and standard deviation for data sampled on cylinder surface with $(r-r_i)/d=0.03$, which is located inside the boundary layer. The PDF is positively skewed since the plumes with higher speed dominate near the inner cylinder \citep{dong2007direct}. Both tails of the PDFs shrink with the increase of polymer concentration, as already observed in the PDFs of $r^3 u^\prime_r \omega$. The reduction in higher angular velocity fluctuations implies that polymers steady the velocity boundary layer.

Similar to the hot and cold plumes defined in Rayleigh-B\'{e}nard turbulence \citep{xie2015effects}, we define fast and slow plumes for TC turbulence. Specifically, they are defined as a time period when $\pm\left[ \omega-\left\langle \omega \right\rangle_t \right]>b\sigma_{\omega}$, with $b=3$, `$+$' for fast plumes and `$-$' for slow ones. The plume fraction $F_p$ is the probability of observing these plumes in a unit time. Plume fraction normalized by its Newtonian value, i.e. $F_p/F_p(\phi=0)$, is shown in Fig. \ref{fig:pdf_uomega}(\textit{b}). For slow plumes, $F_p/F_p(\phi=0)$ drops sharply with the addition of polymers; while for fast plumes, $F_p/F_p(\phi=0)$ drops when $\phi>4$ ppm. Polymers steady the velocity boundary layer, which results in a reduction of both the plume emission rate and the angular velocity flux (Fig. \ref{fig:pdf_nu}(\textit{a})). The reduction in the number of plumes detached from the boundary layer also feeds less energy to the bulk, leading to lower turbulent fluctuations (Fig. \ref{fig:rs}(\textit{b})). \textcolor{black}{We tried different values of $b$ from 2 to 3.5 and the normalized plume fraction based on different values of $b$ shows similar trend as Fig. \ref{fig:pdf_uomega}(\textit{b}).}

\subsection{Velocity power spectra}

\begin{figure}
	\centerline
	{\includegraphics[width=0.8\columnwidth]{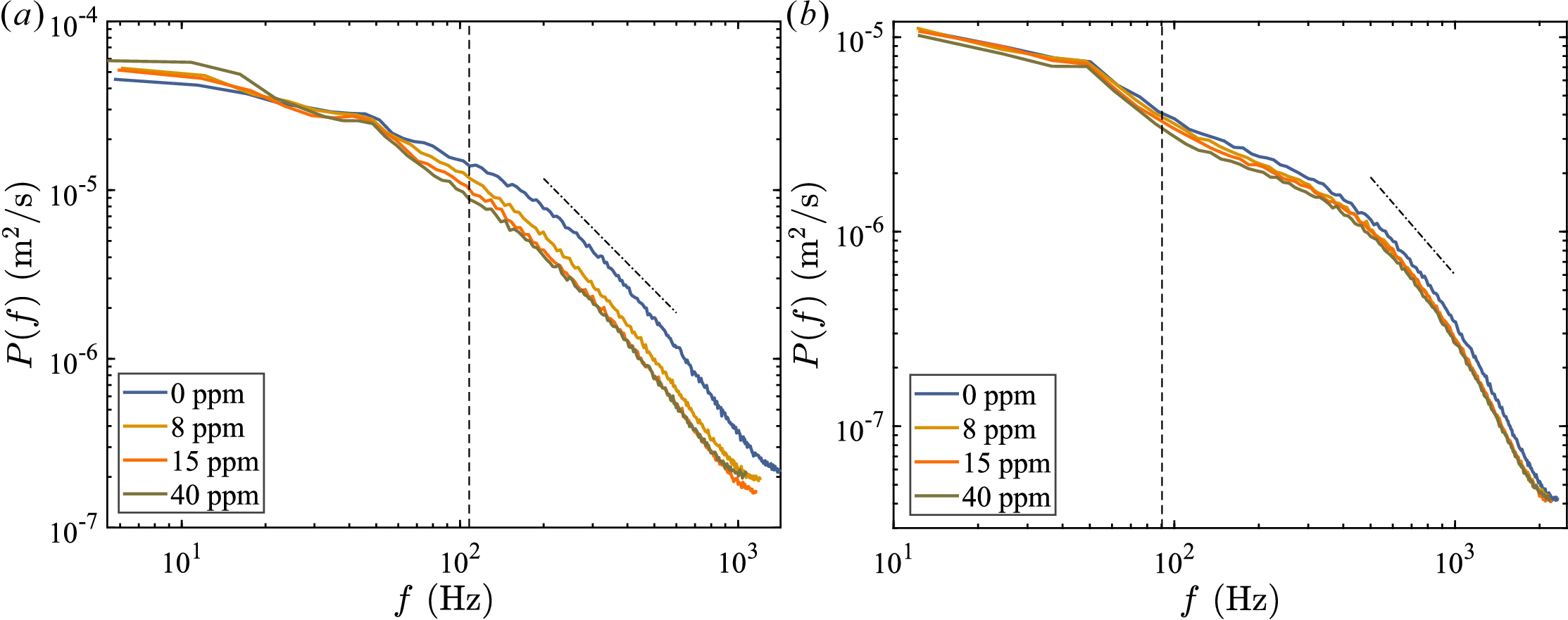}}
	\caption{ Power spectra for the azimuthal velocity $u^\prime_\theta$ at $(r-r_i)/d=0.06$ (\textit{a}) and $(r-r_i)/d=0.5$ (\textit{b}). The vertical dashed lines denote $\left\langle u_\theta \right\rangle_t/d$, which is the characteristic frequency corresponding to the gap width. The dashed-dotted lines have a slope of $-5/3$. $Re=1.3\times 10^4$.}
	\label{fig:psd}
\end{figure}

\textcolor{black}{In the previous section}, we have shown that polymers reduce the turbulence fluctuations (Fig. \ref{fig:rs}(\textit{b})). Turbulent flows are characterized by fluctuations with a wide spectrum of length and time scales \citep{Pope2000}. The effect of polymers on the different scales of turbulence is investigated by the kinetic energy spectrum of the azimuthal velocity $u_\theta^\prime$. The time series of LDA measurements are linearly interpolated using twice the average acquisition frequency to create a time series with equal temporal spacing \citep{huisman2013statistics}. Discrete Fourier transformation is applied to the time series to obtain the spectrum, which is shown in Fig. \ref{fig:psd}(\textit{a}) near the inner cylinder ($(r-r_i)/d=0.06$) and in Fig. \ref{fig:psd}(\textit{b}) at the middle gap ($(r-r_i)/d=0.5$). In Fig. \ref{fig:psd}, we also plot the frequency corresponding to the gap width $\left\langle u_\theta \right\rangle_t/d$ for the $\phi=40$ ppm case as a reference. Near the inner cylinder, the energy content of turbulent fluctuations with scales smaller than $d$ are strongly depressed, \textcolor{black}{which is a result of the suppression of the non-linear energy transfer mechanism by polymers \citep{Ouellette2009,Xi2013,Zhang2021}}. In the near-wall region of TC turbulence, these small-scale structures are G\"{o}rtler vortices, providing direct evidence to our observation in Fig. \ref{fig:rs}(\textit{a}) that G\"{o}rtler vortices are suppressed by polymers. For lower frequency, see $f<20$ Hz or scale larger than $5d$ (here we invoke the Taylor-frozen hypothesis), the energy content is slightly enhanced. \textcolor{black}{These large-scale structures are azimuthal velocity streaks \citep{dong2007direct}. Polymers suppress the non-linear energy transfer from large to small-scales \citep{Ouellette2009,Xi2013,Valente2014,Zhang2021}, and the injected energy may accumulate in these large-scale structures.} Thus, the effect of polymers on the turbulent fluctuations is a redistribution of energy among different scales: suppression of small-scale random fluctuation and energization of \textcolor{black}{large-scale azimuthal velocity streaks}. Since the former effect dominates, the overall effect of polymers is to suppress turbulence fluctuation in the near wall region as shown in Fig. \ref{fig:rs}(\textit{b}).

\begin{figure}
	\centerline
	{\includegraphics[width=0.8\columnwidth]{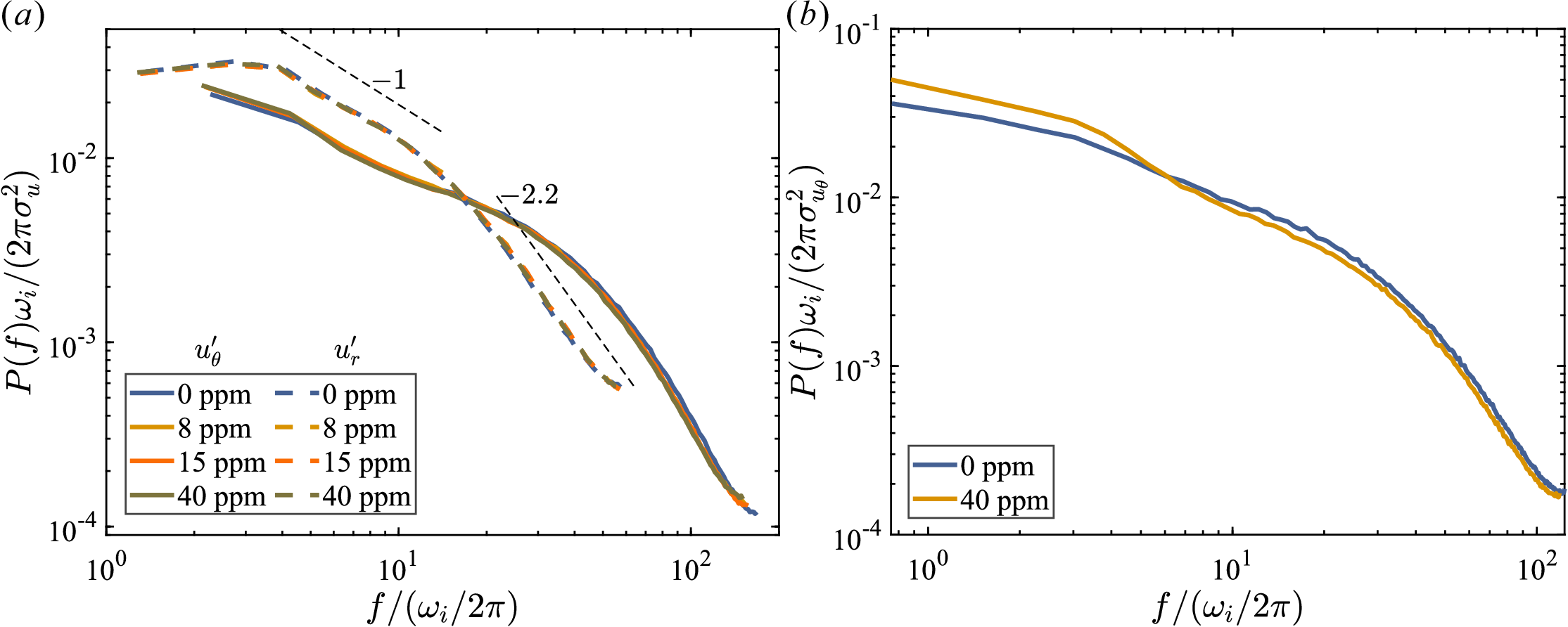}}
	\caption{(\textit{a}) Power spectra normalized by $\omega_i/(2\pi)$ and rms velocity ($\sigma_{u_\theta}$ for azimuthal velocity $u^\prime_\theta$ (solid lines) and $\sigma_{u_r}$ for radial velocity $u^\prime_r$ (dashed lines)). Data shown here are from $Re=1.3\times 10^4$. (\textit{b}) Same as (\textit{a}) but only for $u^\prime_\theta$ at $Re=1.8\times 10^4$. The LDA measurement position is at $(r-r_i)/d=0.5$.}
	\label{fig:psd_norm}
\end{figure}

In wall-bounded turbulence, the velocity gradient is much higher near the wall than in the centre. The ability of turbulent flow to stretch polymer molecules at middle gap can be more precisely quantified by the local Weissenberg number $Wi_{\rm local}$, defined as $Wi_{\rm local}=t_p/t_\eta$, where $t_\eta=\sqrt{\nu/\epsilon_d}$ is the Kolmogorov time scale at $(r-r_i)/d=0.5$. $\epsilon_d\approx0.1\tau\omega_i/\left[ \pi(r_o^2-r_i^2)L\rho_s \right]$ is the average energy dissipation rate in the bulk \citep{ezeta2018turbulence}. At $Re=1.3\times10^4$, $Wi_{\rm local}\approx2.6$, which is smaller the coil-stretch transition Weissenberg number $Wi_c=3\sim4$ reported in homogeneous isotropic turbulence \citep{watanabe2010coil}. Polymers cannot be stretched locally near the middle of the gap at $Re=1.3\times10^4$. From Fig. \ref{fig:psd}(\textit{b}), we observe that all scales resolved here are inhibited by polymers. This inhibition instead reflects the effect of polymers on the velocity boundary layer: polymers steady the boundary layer and reduce the emission rate of velocity plume detached from the boundary layer, resulting in less energy feed into the bulk. To prove this, we show in Fig. \ref{fig:psd_norm}(\textit{a}) the normalized power spectra by \textcolor{black}{$\omega_i/(2\pi)$} and the rms velocity ($\sigma_{u_\theta}$ for azimuthal velocity $u^\prime_\theta$ (solid lines) and $\sigma_{u_r}$ for radial velocity $u^\prime_r$ (dashed lines)). \textcolor{black}{The integral of the normalized spectra with respect to $f/(\omega_i/2\pi)$ would be 1.} The normalized spectra nearly overlap with each other, in support of the above discussion. At a higher Reynolds number $Re=1.8\times10^4$ ($Wi_{\rm local}\approx4.1$), all the scales are suppressed by polymers as the $Re=1.3\times10^4$ case. However, the spectra cannot be collapsed by the above normalization (see Fig. \ref{fig:psd_norm}(\textit{b})). Instead, the spectrum of $\phi=40$ ppm changes its slope at an intermediate frequency, above and below which the energy content is increased and reduced respectively, \textcolor{black}{implying that the turbulent structures are altered in the bulk region at higher $Re$ due to the larger $Wi_{\rm local}$.}

It is interesting to note that \textcolor{black}{the spectra of $u^\prime_r$ exhibit two power-law scaling with a $-1$ slope for intermediate frequency and a $-2.2$ slope for high frequency, and the crossover happens at $f_c/(\omega_i/2\pi)) \approx 16$}. This behaviour is reminiscent of what was observed in elastic turbulence \citep{groisman2004elastic}. In elastic turbulence, where the elastic stress dominates the dynamics and inertia can be neglected, \textcolor{black}{\cite{groisman2004elastic} reported that the spectra of $u^\prime_r$ show respectively the $-1.1$ and $-2.2$ power-law decay for intermediate and high frequency in TC flow}. However, in their flow the crossover between the two power-laws happens at $f_c/(\omega_i/2\pi) \approx 0.45\pm0.05$. \textcolor{black}{The similar decay slope suggests that one cannot discriminate whether the flow is dominated by elasticity or inertia based solely on the spectrum of $u^\prime_r$ in TC turbulence.} The crossover frequency is much higher in inertia-dominated flow than the elastic turbulence, indicating that much smaller scale structures are developed in the former case. The reason for the similar scaling observed in our study and in elastic turbulence is not known to us yet.

\textcolor{black}{In elasto-inertial turbulence of TC flow, experiments have reported spatial spectra with decay exponents of $-7/3$ \citep{moazzen2023friction} and $-2.6$ \citep{boulafentis2024coherent}, while numerical simulations have found exponents of $-3$ \citep{lopez2022vortex} or less than $-3$ \citep{song2021direct,song2021reverse}. For the Reynolds numbers investigated here, we do not observe a clear inertial scaling range with an exponent of $-5/3$ for $u^\prime_\theta$ in either the near-wall or bulk regions, as shown in Fig. \ref{fig:psd}, likely due to the Reynolds number not being high enough \citep{huisman2013statistics}. In TC flow with polymers, we also cannot assign a definitive scaling exponent to the spectra of $u^\prime_\theta$ because of the limited scaling range. Further experiments are needed to characterize the spectra in viscoelastic TC flow across different parameter regimes.}

\subsection{Velocity boundary layer}

\begin{figure}
	\centerline
	{\includegraphics[width=0.8\columnwidth]{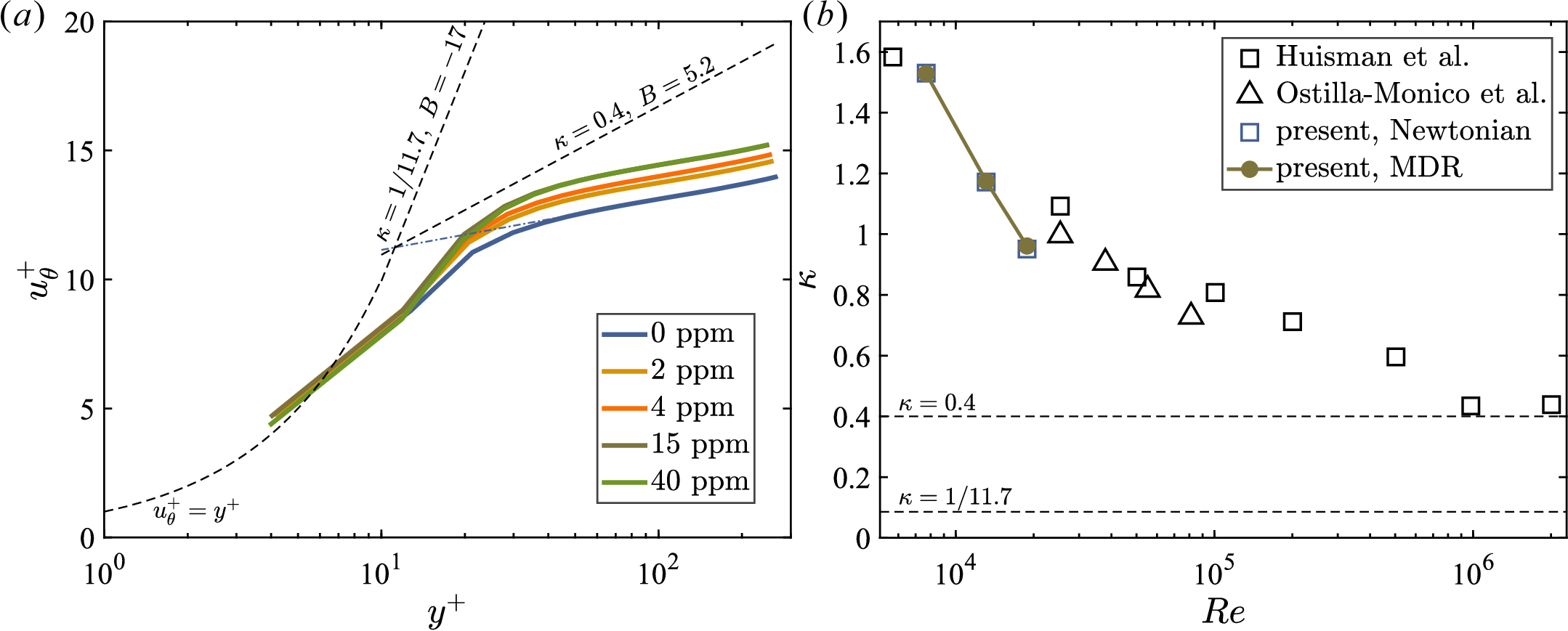}}
	\caption{(\textit{a}) Azimuthal velocity profiles near the inner cylinder for varying polymer concentration at $Re=1.3\times 10^4$. $u^+_\theta=y^+$ is the viscous sublayer profile. The logarithmic law of the wall $u^+_\theta=1/\kappa {\rm ln} y^++B$ with typical values of $\kappa=0.40$ and $B=5.0$ is also shown for comparison. The logarithmic law with values of $\kappa=1/11.7$ and $B=-17$ is the maximum drag reduction asymptote reported in channel flow. (\textit{b}) The parameter $\kappa$ as a function of $Re$ for both Newtonian case and viscoelastic case under the maximum drag reduction condition. $\kappa$ from \cite{huisman2013logarithmic} and \cite{ostilla2014boundary} are also shown.}
	\label{fig:mv}
\end{figure}

\textcolor{black}{In the previous sections}, we discuss the effects of polymers on the Taylor vortex and turbulent fluctuations. The mean velocity profile near the wall is of great interest in the study of polymer drag reduction. \cite{warholic1999influence} divided the effect of polymers into two regimes based on the degree of drag reduction: low drag reduction regime where the mean streamwise velocity profile for the viscoelastic case shifts upward in the logarithmic law layer with its slope unchanged; high drag reduction regime where the slope of the mean streamwise velocity profile increases with drag reduction rate and approaches an empirical asymptote at MDR. The mean azimuthal velocity in TC flow is defined as
\begin{equation}
	u^+_\theta(y^+)=\left[ u_\theta(r_i)-\left\langle u_\theta(r) \right\rangle_{t,z} \right]/u_\tau,
\end{equation}
where $\left\langle u_\theta(r) \right\rangle_{t,z}$ is the azimuthal component of the velocity, and $u_\theta(r_i)=\omega_i r_i$ is the azimuthal velocity of the inner cylinder. $y^+=(r-r_i)/\delta_\nu$ is the distance from the inner cylinder in unit of the viscous length scale $\delta_\nu=\nu/\sqrt{\tau/(2\pi \rho_s r_i^2L)}$.

$u^+_\theta$ as a function of $y^+$ for various polymer concentrations is shown in Fig. \ref{fig:mv}(\textit{a}), along with the $u^+_\theta=y^+$ profile in the viscous sublayer and the logarithmic law $u^+_\theta=1/\kappa {\rm ln} y^++B$. For pipe, flat plate, and channel flows, $\kappa=0.40$ and $B=5.0$ \citep{marusic2013logarithmic}. For channel flow laden with polymers, $\kappa=1/11.7$ and $B=-17$ at MDR \citep{virk1970}. We observe that $u^+_\theta$ is lower than the logarithmic law of Newtonian turbulence in flat plate boundary layer, but it still exhibits a logarithmic shape for $y^+>50$. The growth rate of $u^+_\theta$ in the logarithmic layer is smaller than $\kappa=0.40$. This is not surprising since the value $\kappa=0.40$ was obtained for a zero pressure gradient boundary layer, which is not satisfied in TC turbulence \citep{huisman2013logarithmic,ostilla2014boundary}. With the increase of polymer concentration, $u^+_\theta$ shift upward, and their slopes are nearly the same as that of the Newtonian case. We stress that even at MDR the slope of $u^+_\theta$ is not changed by polymers.

We extract the slope of $u^+_\theta$ at three different $Re$. Least-square fitting is applied to $u^+_\theta$ for $y^+>50$ as suggested in \cite{huisman2013logarithmic}. $\kappa$ for both Newtonian case and viscoelastic case at MDR are shown in Fig. \ref{fig:mv}(\textit{b}), where the experimental results from \cite{huisman2013logarithmic} and numerical results from \cite{ostilla2014boundary} are also shown for comparison. $\kappa$ decreases with $Re$, and is in good agreement with previous results. $\kappa$ for Newtonian case and viscoelastic case at MDR are nearly overlapping, suggesting that polymers do not alter the slope of the mean velocity in the current parameter range of TC turbulence.

\section{\textcolor{black}{Conclusions}}\label{sec:con}
We present an experimental study on polymer drag reduction in TC turbulence within the Reynolds number range of $4 \times 10^3 < Re < 2.5 \times 10^4$. Two polyacrylamide polymers with different molecular weights are used with the corresponding elastic number smaller than $0.01$. The flow thus is inertia-dominated. \textcolor{black}{The system drag is measured by a rheometer with high precision.} The local flow field is measured by planar PIV and LDA.

It is found that the drag reduction rate increases with polymer concentration and approaches the MDR limit. At MDR, the friction factor follows a power-law scaling with an effective exponent $-0.58$, i.e., $C_f \sim Re^{-0.58}$, which is close to the exponent reported in channel/pipe flows \citep{virk1970}. However, the drag reduction rate in TC turbulence is about $20\%$, which is much lower than that in rectilinear flows at similar Reynolds numbers. Besides, the Reynolds shear stress does not vanish, and the velocity profile only shifts upward with its slope in the logarithmic layer unchanged at MDR. These behaviours are reminiscent of the low drag reduction regime reported in channel flow \citep{warholic1999influence}.

By separating the advective angular velocity flux into the mean flow (Taylor vortex) and turbulent flow (Reynolds shear stress) contributions, we show that the turbulent contribution is strongly reduced while the mean flow contribution is only slightly reduced. In the current parameter range of TC turbulence, mean flow contribution accounts for more than $50\%$ of the total angular velocity flux in the bulk \citep{brauckmann2013direct}. The lower drag reduction observed in TC turbulence thus originates from the observation that the major effect of polymers is on the turbulent flow and the mean flow is slightly affected. We further reveal that the above results can be traced to the fact that polymers steady the velocity boundary layer, and reduce the emission rate of both fast and slow plumes. The reduced number of intense plumes detached from the boundary layer in turn feeds less energy into the bulk turbulence, resulting in weaker turbulent fluctuations and Reynolds shear stress. In the near-wall region, polymers also redistribute the kinetic energy among different scales: the small-scale G\"{o}rtler vortices are highly suppressed and the \textcolor{black}{large-scale azimuthal velocity streaks} become slightly more energetic.

At high Reynolds numbers, the effects of polymers are predominantly observed at the small-scales of turbulence, as reported in previous studies \citep{Ouellette2009,Perlekar2010,Xi2013,Zhang2021}. We here show that the effects of polymers on the large-scale secondary flow structures is only marginal. When secondary flow structures are statistically persistent and dominate the global transport properties of the system, the drag reduction efficiency of polymers is diminished. \textcolor{black}{At higher Reynolds numbers, the effects of Taylor vortex diminish and turbulent plumes dominate \citep{ostilla2014boundary,grossmann2016}, we would expect that the effects of polymers and the drag reduction rate will become more pronounced. However, whether the vanishing Reynolds shear stress and asymptotic velocity profile can be observed at MDR remains unclear.} The current work represents an advancement by identifying the similarities and differences in drag reduction mechanisms across different turbulence flow systems, setting the stage for further research into turbulent flows with polymer additives at high Reynolds number regimes in TC turbulence.

\backsection[Acknowledgements]{This work is financially supported by the National Natural Science Foundation of China under Grant Nos. 11988102, 12125204, 12388101, 12402298, and 12402299, the New Cornerstone Science Foundation through the New Cornerstone Investigator Program, the XPLORER PRIZE, and the 111 Project of China through Grant No. B17037.}

\backsection[Declaration of Interests]{The authors report no conflict of interest.}

\backsection[Author ORCIDs]{\\Yi-Bao Zhang https://orcid.org/0000-0002-4819-0558; \\
Yaning Fan https://orcid.org/0009-0000-5886-3544;\\
Jinghong Su https://orcid.org/0000-0003-1104-6015;\\
	Heng-Dong Xi https://orcid.org/0000-0002-2999-2694; \\
	Chao Sun https://orcid.org/0000-0002-0930-6343.}

\appendix
\section{Viscosity measurement}\label{appA}

\begin{figure}
	\centerline
	{\includegraphics[width=0.8\columnwidth]{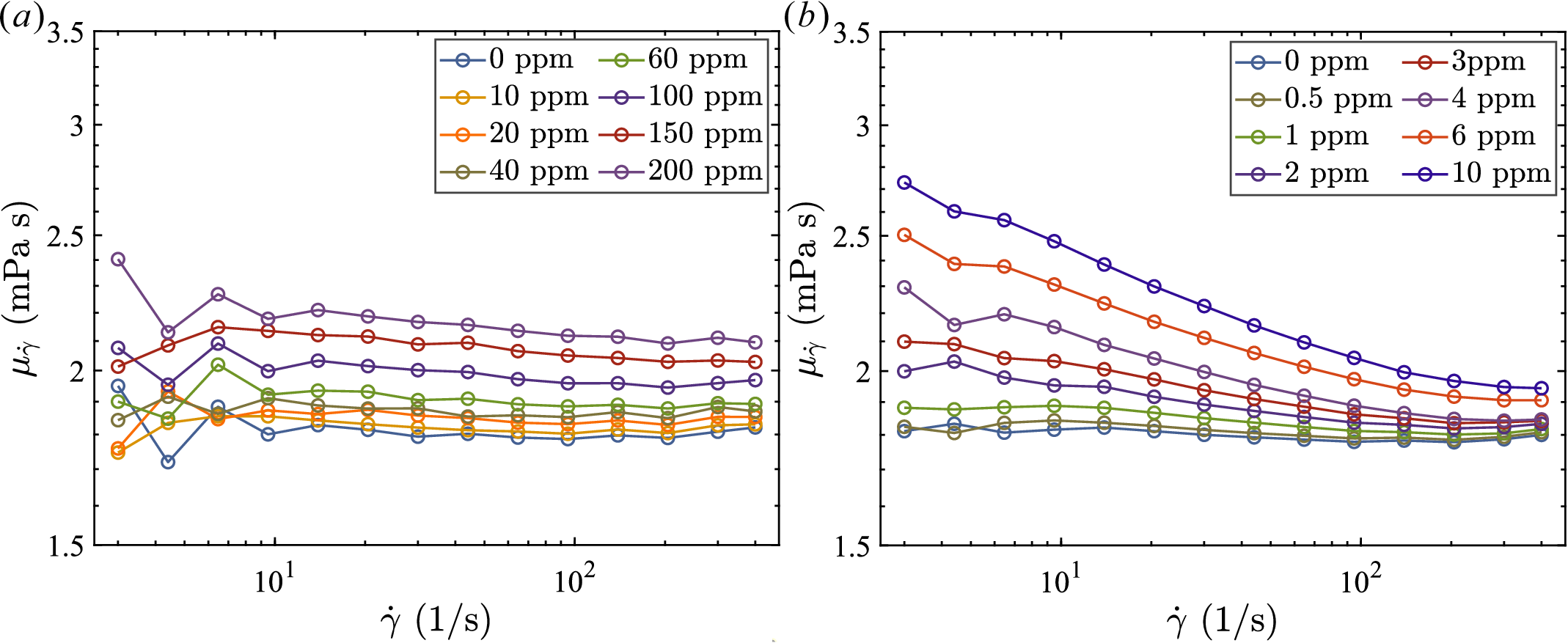}}
	\caption{Shear viscosity $\mu_{\dot{\gamma}}$ as a function of shear rate $\dot{\gamma}$ for pam5e6 (\textit{a}) and pam2e7 (\textit{b}).}
	\label{fig:visco}
\end{figure}

The shear viscosity of the polymer solutions is measured by the Discovery Hybrid Rheometer equipped with a cone-plate geometry, whose diameter is 40 mm and angle is $2^\circ$. \textcolor{black}{For pam5e6, the shear viscosity $\mu_{\dot{\gamma}}$ is nearly constant and does not show the shear-thinning effect over the shear rate range of $3<\dot{\gamma}<400$. For pam2e7, $\mu_{\dot{\gamma}}$ slightly decreases with $\dot{\gamma}$ and then approaches a constant when $\dot{\gamma}>200$. To evaluate the extent of shear-thinning of the working fluids, we use the average gradient of the viscosity curve \cite{boulafentis2024coherent},}
\begin{equation}
    n_\mu=\langle \frac{\partial {\rm log}(\mu_{\dot{\gamma}})}{\partial {\rm log}(\dot{\gamma})} \rangle_{\dot{\gamma}} +1,
\end{equation}
\textcolor{black}{whose value is less than or equal to 1. The smaller the value of $n_\mu$, the more significant the shear-thinning effect of the working fluid. In TC flow, the average shear rate in the gap is $\omega_ir_i/d$, which is larger than 80 s$^{-1}$ in this study. $n_\mu$ is calculated from the viscosity curve in Fig. \ref{fig:visco} over $80<\dot{\gamma}<400$. For pam5e6, $n_\mu>0.992$ at $\phi \leq 100$ ppm. For pam2e7, $n_\mu>0.978$ at $\phi \leq 4$ ppm. At these values of $n_\mu$, the shear-thinning effect can be practically neglected, as in many previous studies \citep{boulafentis2024coherent,lacassagne2020vortex}. We thus average the measured shear viscosity over the range of $80<\dot{\gamma}<400$ to obtain an averaged viscosity $\mu$, i.e. $\mu=\langle \mu_{\dot{\gamma}} \rangle _{80<\dot{\gamma}<400}$. This value is used as the viscosity of the working fluid in this study.}

\textcolor{black}{From the viscosity curves, the zero-shear viscosity $\mu_0$ can be obtained \citep{boulafentis2024coherent}. For both polymers, $\mu_0$ increases with polymer concentration, and can be fitted with a second order polynomial, i.e. $(\mu_0-\mu_s)/\mu_s=\left[ \mu \right]\cdot \phi +A\cdot \phi^2$, where $\mu_s$ is the viscosity of the solvent, $\left[ \mu \right]$ the intrinsic viscosity, and $A$ a fitting parameter. The overlap concentration $\phi^*$ is estimated from $\left[ \mu \right]$ by the relation $\phi^*=0.77/\left[ \mu \right]$ \citep{boulafentis2024coherent}. $\phi^*=680$ and 17 ppm for pam5e6 and pam2e7, respectively.}

\section{Dimensionless torque}\label{appB}

\begin{figure}
	\centerline
	{\includegraphics[width=0.5\columnwidth]{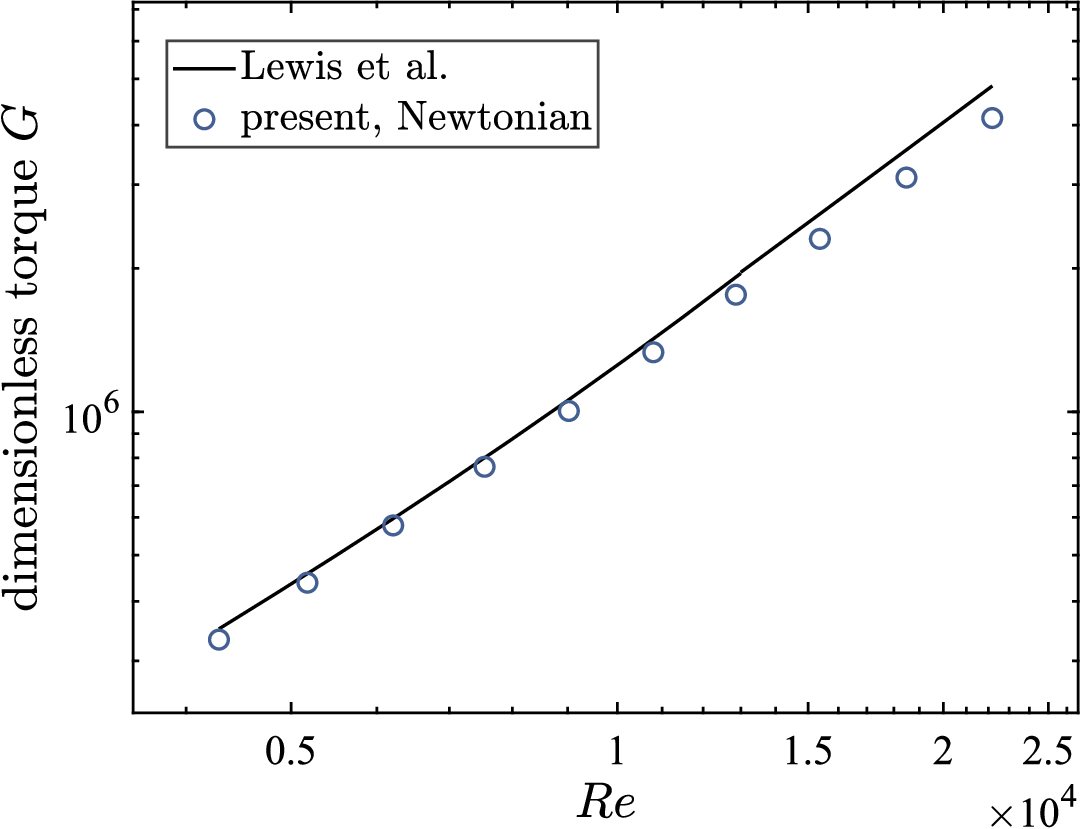}}
	\caption{\textcolor{black}{Comparison of the dimensionless torque, $G$, from the present study in the Newtonian case and Lewis $\&$ Swinney \citep{lewis1999velocity}.}}
	\label{fig:torque_B}
\end{figure}

\textcolor{black}{The dimensionless torque, $G$, measured in this study and from \cite{lewis1999velocity} are compared in Fig. \ref{fig:torque_B}. $G$ in the present study is slightly lower than that in \cite{lewis1999velocity}, which can be attributed to the different radius ratio: $r_i/r_o=0.714$ in this study and $r_i/r_o=0.724$ in \cite{lewis1999velocity}.}

\bibliographystyle{jfm}
\bibliography{polymer}

\end{document}